\documentclass[authoryear,3p,review]{elsarticle}

\usepackage{amssymb}

\usepackage{url}
\usepackage{color, colortbl}
\usepackage{graphicx}

\usepackage{xcolor}
\newcommand{\revised}[1]{#1}
\usepackage{lineno}

\journal{Planetary and Space Sciences}

\begin{document} 

\begin{frontmatter}

   \title{Jovian Auroral Radio Source Occultation Modelling and Application 
   to the JUICE Science Mission Planning}
   

   \author[1]{Baptiste Cecconi\corref{cor1}}
   \ead{baptiste.cecconi@observatoiredeparis.psl.eu}

   \author[2]{Corentin K. Louis}
   \ead{corentin.louis@dias.ie}
   
   \author[3]{Claudio Mu\~noz Crego}
   \ead{cmunoz@sciops.esa.int}
   
   \author[4]{Claire Vallat} 
   \ead{cvallat@sciops.esa.int}
   
   \cortext[cor1]{Corresponding author}
    
   \affiliation[1]{
      organization={LESIA, Observatoire de Paris, CNRS, PSL Research University},
      city={Meudon}, 
      country={France}}

   \affiliation[2]{
      organization={School of Cosmic Physics, DIAS Dunsink Observatory, Dublin 
      Institute for Advanced Studies}, 
      city={Dublin}, 
      country={Ireland}}
         
   \affiliation[3]{
      organization={Aurora B.V., for European Space Agency, ESAC}, 
      city={Madrid}, 
      country={Spain}}

   \affiliation[4]{
      organization={Rhea Group, for European Space Agency, ESAC}, 
      city={Madrid}, 
      country={Spain}}
 
\begin{abstract}
Occultations of the Jovian low frequency radio emissions by the Galilean moons have 
been observed by the PWS \revised{(Plasma Wave Science)} instrument of the Galileo 
spacecraft. We show that the ExPRES (Exoplanetary and Planetary Radio Emission 
Simulator) code accurately models the temporal occurrence of the occultations in the 
whole spectral range observed by Galileo/PWS. This validates of the ExPRES code. 
\revised{In addition to supporting the analysis of the science observations, t}he 
method can be applied for preparing the JUICE moon flyby science operation planning.
\end{abstract}


\end{frontmatter}
%

\section{Introduction} 

The magnetosphere of Jupiter produces low frequency radio emissions \revised{near 
the planetary polar} regions, along the active magnetic field lines connected to 
the Jovian auroral oval as well as to the Galilean moon auroral magnetic footprints.  
The Jovian radio emissions are intense and non-thermal radio frequency phenomena, 
spanning from a few kHz to about 40 MHz\revised{. They} are produced through the 
Cyclotron Maser Instability (CMI), which converts the local plasma free-energy 
into electromagnetic radiation \revised{\citep{Zarka:2004hk,Louarn:2017bc,Louarn:2018cb}}. 
They are used as a proxy for the Jovian magnetospheric activity.  They \revised{were} 
discovered by \citet{burke_JGR_55} and have been since studied with ground 
observatories \citep[e.g., Nan\c cay \revised{Decametre} Array,][]{2017pre8.conf..455L} 
and space-borne instruments (with, e.g., the Voyager, Galileo, Cassini and Juno 
space missions). 

\begin{figure*}
    \includegraphics[width=\linewidth]{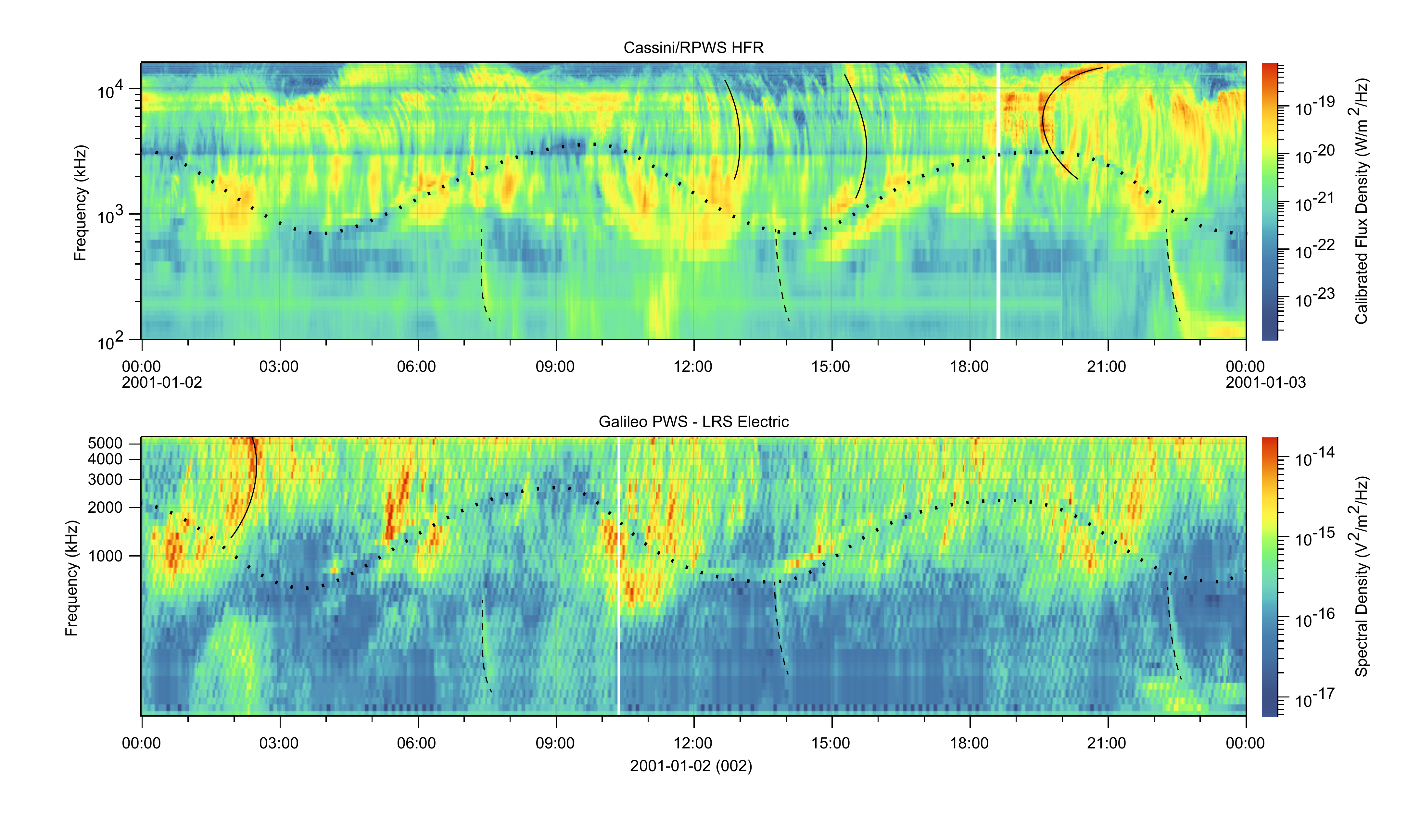}
    \caption{Calibrated Cassini/RPWS/HFR \citep{zarka_JGR_04,https://doi.org/10.25935/h98j-ma66} 
    (top) and GLL/PWS (see Section \ref{gll-pws-data}) (bottom) radio electric power spectral 
    densities during 24 hours close to the Cassini flyby of Jupiter. Significant features have 
    been highlighted: a few radio emission ``arcs'' are  traced in plain line; Type III Solar 
    radio bursts are traced in dashed line; and the attenuation lanes are traced in dotted 
    line. }\label{fig:cas-gll-continuous-radio}
\end{figure*}

Observations from the Cassini Radio and Plasma \revised{Wave} Science 
\citep[RPWS,][]{gurnett_SSR_04} and Galileo Plasma \revised{Wave} Science 
\citep[PWS,][]{gurnett_SSR_92} experiments showed that the Jovian radio emission events 
are observed quasi-permanently along the spacecraft orbit. Figure 
\ref{fig:cas-gll-continuous-radio} shows 24 hours of simultaneously observed spectral 
flux densities by the Cassini/RPWS and Galileo/PWS instruments, when Cassini was close to 
its flyby of Jupiter. Both panels of this figure display many arc-shaped radio events (a 
few of them are highlighted with a thin plain black line), with a few small time-frequency 
regions with quiet (background level) conditions. Radio arcs can be classified with 
orientation of their curvature: ``vertex-early'' and ``vertex-late'' arcs corresponds to 
opening ``('' or closing ``)'' parenthesis shapes. The arc shape of Jovian radio emission 
is well explained by the CMI mechanism at the radio source, coupled with the shape of the 
magnetic field lines, and the rotation of Jupiter with respect of the observer, as 
described in Fig.\ 1 of \citet{Louis:2019gp}. This figure also \revised{introduces} two 
other features of the observed radio spectrum around Jupiter: Type \revised{III} Solar 
radio bursts are also observed depending on the solar activity (three events are 
highlighted with a thin dashed black line); and the so-called ``attenuation lanes''
\citep{gurnett_GRL_98a} resulting from the propagation of hectometric waves through the 
Io plasma torus \citep{menietti_PSS_03}. The attenuation lanes are observed as a 
narrow-band attenuation feature modulated at the planetary rotation period. The 
attenuation is also accompanied or replaced with an intensification of the signal, 
\revised{similar} to caustic optical phenomena. \revised{The Solar Type III bursts are 
the only features observed simultaneously on each panel. This illustrates the strong
anisotropy of the radio emission features at Jupiter, since the two spacecraft are located
at different longitude around the planet.}

Although not covering the full spectral range of the Jovian radio emissions, Galileo/PWS 
\revised{routinely collected} radio observations during its many orbits in the Jovian 
system. This data \citep{PWS_LPW_PDS} shows quasi continuous emission from a few 100 kHz 
up to 5.6 MHz, the upper spectral limit of PWS. During the Galilean moon flybys, 
Galileo/PWS observed full occultations of the Jovian radio emissions \citep{Kurth:1997in}.

The ESA JUICE \citep[JUpiter ICy moon Explorer,][]{2019EPSC...13..400W} will explore the 
Jupiter system and its magnetosphere. The study of the Jovian magnetosphere can strongly 
benefit from remote observations and modelling tools \citep{cecconi_baptiste_2019_2583611}. 
Two instruments of the JUICE scientific payload \revised{will operate} in the low frequency 
radio range (below 50 MHz): the Radio and Plasma Waves Instrument \citep[RPWI, 
][]{2013EPSC....8..637W} has a receiver dedicated to the study of Jovian radio emissions; 
and the Radar for Icy Moon Exploration (RIME) experiment \citep{Bruzzone:2013ge} 
\revised{will operate} with a central frequency at 9\,MHz, which lies within the Jovian 
radio emission spectral range. The Jovian radio emission may interfere with RIME active 
radar mode \citep{cecconi_PSS_11}, but can be also used in a passive radar experiment mode 
during icy moon flybys \citep{RomeroWolf:2015fja,Schroeder:2016bu,2017pre8.conf..127K}.

The ExPRES code \citep[Exoplanetary and Planetary Radio Emissions Simulator,][]{Louis:2019gp} 
simulates for a given observer the geometrical visibility of radio emissions of a 
magnetised body. This visibility depends in particular of \revised{the angle} between the 
magnetic field vector at the source and the emitted wave vector, which is computed 
self-consistently in the frame of the CMI theory. The anisotropic shape of the radio source 
and their geometrical observability conditions are well described in Fig.\ 1 (panels c, d 
and e) of \citet{Louis:2019gp}. This computation is iterated at each time/frequency step 
and for each source. The produced time-frequency map (or dynamic spectrum) can then 
directly be compared to observations. \revised{ExPRES thus provides a characterization 
(time, frequency, location, polarization, etc.) of the observable auroral radio sources 
of Jupiter for an observer at a given location. It can be used by any team or processing 
requiring such kind of information.}

In this study, we \revised{model} the Jovian radio emission occultations during (past) 
Galileo and (planned) JUICE Galilean moon flybys, using the ExPRES code. \revised{We use 
the ExPRES model results to interpret the Galileo/PWS observed occultations, and show how 
this model can be used to prepare the JUICE mission science operation planning. We also 
discuss how the low frequency radio occultation can be used to characterise the Galilean 
moon's environment.}


\section{Data Sets} 
Several sets of data have been used in this study and are presented in this section. 
For the Galileo spacecraft flybys, we compare the actual observed radio \revised{spectra 
with} low frequency radio occultations predicted by the ExPRES code, using the actual 
flyby geometry (spacecraft and moon trajectory in a Jovian reference frame). The JUICE 
spacecraft study only includes modelled \revised{occultations.}

\subsection{Galileo PWS Observations}
\label{gll-pws-data}
All Galilean moon flyby of Galileo with PWS data have been modelled and analysed. The 
Galileo PWS (hereafter referred to as GLL/PWS) data have been retrieved from \revised{the} 
University of Iowa \emph{das2} server interface \citep{piker_2019}, using the 
\texttt{das2py}\footnote{Available from \url{https://github.com/das-developers/das2py} 
(last access: 17-Aug-2021)} python module. We have used the \emph{GLL/PWS LRS 152-channel 
calibratedelectric} collection\footnote{URI:
\url{http://das2.org/browse?resolve=tag:das2.org,2012:site:/uiowa/galileo/pws/survey_electric}} 
from that \emph{das2} server. The data are radio-electric power spectral densities 
provided in units of V$^2$/m$^2$/Hz. This dataset doesn't include the instrument's 
antenna gain calibration, but this has no consequence on this study. The data set has 
a native time resolution of 18.67 seconds. These data are also available in the full 
resolution GLL/PWS dataset \citep[\texttt{GO-J-PWS-2-REDR-LPW-SA-FULL-V1.0}, 
][]{PWS_LPW_PDS}, at NASA PDS (Planetary Data System) PPI (Planetary Plasma Interaction) 
node. 

\subsection{Moons and Spacecraft Trajectory Data}
The moons and spacecraft trajectory data are computed using SPICE kernels 
\citep{1996P&SS...44...65A}. In this study, the ephemeris data have been retrieved using 
the NASA-JPL (Jet Propulsion Laboratory of the National Aeronautics and Space Administration) 
instance of WebGeoCalc \citep{Acton:2018dd} for \revised{the} Galileo spacecraft and Jovian 
moons, and another WebGeoCalc instance at ESA-ESAC (European Space Astronomy Centre of the 
European Space Agency) for the JUICE spacecraft. The JUICE spacecraft SPICE kernels
\citep{juice-kernet-set} contains all the studied orbital scenarii, as described in the JUICE 
CReMA (Consolidated Report on Mission Analysis) documents. In this study, the selected JUICE 
trajectory scenario is \revised{CReMA version 3.0 (hereafter notes CReMA-3.0)\footnote{CReMA-3.0 
trajectory information: \url{https://www.cosmos.esa.int/web/juice/crema-3.0} (last access: 
17-Aug-2021)}.} The ephemeris of all bodies have been retrieved in the \verb|IAU_JUPITER| 
reference frame, also referred to as ``IAU Jupiter System III (1965)''. In the WebGeoCalc 
interface, we use the ``planetocentric'' representation for coordinate retrieval, in which 
the longitude is oriented Eastward. 

For each flyby, two ephemeris data files are retrieved, using the `State Vector' 
WebGeoCalc capability: (a) the location of the moon and (b) that of the spacecraft, both 
in the \texttt{IAU\textunderscore JUPITER} frame, as seen from the center of Jupiter, 
with a time interval of a few hours (2 to 4 hours, depending on the spacecraft velocity 
relative to the moon) \revised{centered} on the closest approach epoch of the flyby, and a 
time sampling step of one minute. We do not correct for light time propagation. The resulting 
uncertainty in ephemeris data timing is of the order of 1 second, which is much below 
time resolution of the data and the simulations.

\section{Jovian Radio Emissions Occultations}

\begin{figure}[ht]
    \centering
    \includegraphics[width=\linewidth]{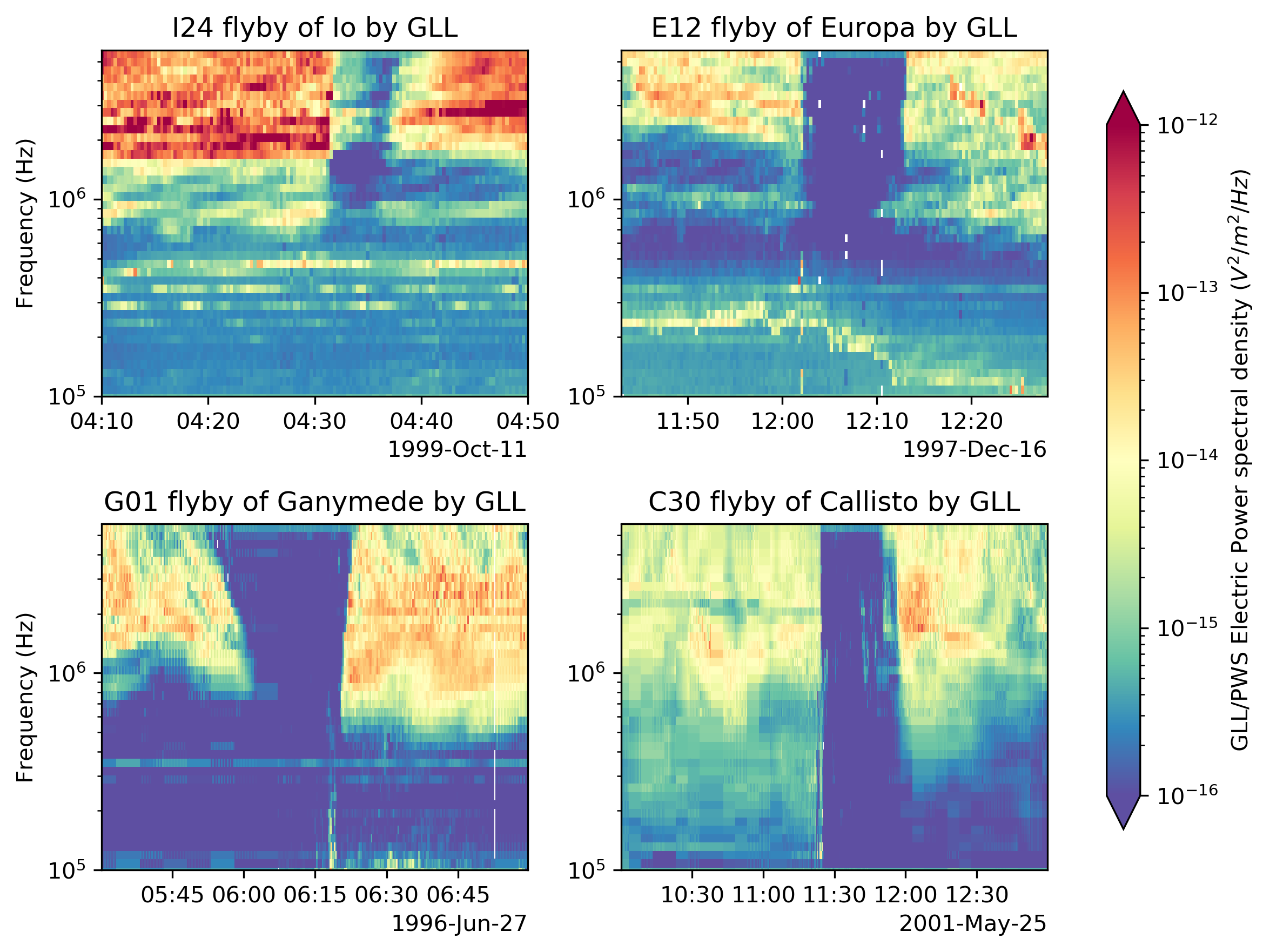}
    \caption{Jovian radio emission occultations by Io (upper left), Europa (upper 
    right), Ganymede (lower left) and Callisto (lower right), as seen by GLL/PWS. The 
    figures are showing radio electric power spectral densities in V$^2$/m$^2$/Hz.}
    \label{fig:flyby_data}
\end{figure}

As shown by \citet{Kurth:1997in}, the Jovian hectometric radio emissions are occulted by 
Ganymede during \revised{the} G01 flyby (Ganymede flyby during the first orbit around 
Jupiter) of the Galileo spacecraft, on June 27th 1996. Figure 1 of \citet{Kurth:1997in} 
shows the GLL/PWS spectrogram during G01 flyby (also displayed on the bottom-left panel 
of Figure \ref{fig:flyby_data}). The full occultation is observed between 05:50 and 06:20 
SCET. The occultation spectral ingress and egress profiles imply that the observed radio 
sources at higher frequencies (located close to Jupiter) are occulted earlier and reappears 
later than the lower frequency ones, which are located \revised{farther} out from Jupiter. 
Possible occultation of the Jovian radio emissions have been also reported during the first 
Io flyby of Galileo \citep{louarn_GRL_97}. 

Table \ref{tab:gll-flyby-list} shows our assessment of all targeted Galilean moon flybys 
by the Galileo spacecraft. Appendix \ref{app:supplementary} provides access to the full 
material used to conduct this study, with figures corresponding to each flyby. In this 
paper, we have selected one flyby of each Galilean moon, where the radio emission 
occultation was clearly observed (see the grey rows in Table \ref{tab:gll-flyby-list}). 
Figure \ref{fig:flyby_data} shows GLL/PWS observations for each of the selected flybys, 
i.e., from left to right and top to bottom: Io (I24), Europa (E12), Ganymede (G01) and 
Callisto (C30) flybys.

\begin{table*}[ht]
    \centering
    \begin{tabular}{c|c|c|c|c|c}

\hline 
Orbit & Moon &        Moon      &  \multicolumn{2}{c|}{Data Availability} &  Occultation \\
Name  & Name & Closest Approach &  GLL/PWS & SPICE &  Assessment \\
\hline
I00 & Io       & \texttt{1995-12-07 17:45:58} & yes & yes & \emph{(?)} \\
\rowcolor{lightgray}
G01 & Ganymede & \texttt{1996-06-27 06:29:07} & yes & yes & \emph{yes} \\
G02 & Ganymede & \texttt{1996-09-06 18:59:34} & yes & yes & \emph{(?)} \\
C03 & Callisto & \texttt{1996-11-04 13:34:28} & yes & yes & \emph{(?)} \\
E04 & Europa   & \texttt{1996-12-19 06:52:58} & yes & yes & \emph{no } \\
E06 & Europa   & \texttt{1997-02-20 17:06:10} & yes & yes & \emph{(?)} \\
G07 & Ganymede & \texttt{1997-04-05 07:09:58} & yes & yes & \emph{(?)} \\
G08 & Ganymede & \texttt{1997-05-07 15:56:10} & yes & yes & \emph{yes} \\
C09 & Callisto & \texttt{1997-06-25 13:47:50} & yes & yes & \emph{(?)} \\
C10 & Callisto & \texttt{1997-09-17 00:18:55} & yes & yes & \emph{no } \\
E11 & Europa   & \texttt{1997-11-06 20:31:44} & yes & yes & \emph{no } \\
\rowcolor{lightgray}
E12 & Europa   & \texttt{1997-12-16 12:03:20} & yes & yes & \emph{yes} \\
E14 & Europa   & \texttt{1998-03-29 13:21:05} & yes & yes & \emph{(?)} \\
E15 & Europa   & \texttt{1998-05-31 21:12:57} & yes & no  & \emph{yes} \\
E16 & Europa   & \texttt{1998-07-21 05:03:45} & yes & yes & \emph{(?)} \\
E17 & Europa   & \texttt{1998-09-26 03:54:20} & yes & yes & \emph{(?)} \\
E18 & Europa   & \texttt{1998-11-22 11:38:26} & no  & yes & \emph{---} \\
E19 & Europa   & \texttt{1999-02-01 02:19:50} & yes & yes & \emph{(?)} \\
C20 & Callisto & \texttt{1999-05-05 13:56:18} & yes & yes & \emph{no } \\
C21 & Callisto & \texttt{1999-06-30 07:46:50} & yes & yes & \emph{no } \\
C22 & Callisto & \texttt{1999-08-14 08:30:52} & yes & yes & \emph{yes} \\
C23 & Callisto & \texttt{1999-09-16 17:27:02} & yes & yes & \emph{yes} \\
\rowcolor{lightgray}
I24 & Io       & \texttt{1999-10-11 04:33:03} & yes & yes & \emph{yes} \\
I25 & Io       & \texttt{1999-11-26 03:59:15} & yes & yes & \emph{(?)} \\
E26 & Europa   & \texttt{2000-01-03 17:59:56} & yes & yes & \emph{no } \\
I27 & Io       & \texttt{2000-02-22 13:46:36} & yes & yes & \emph{yes} \\
G28 & Ganymede & \texttt{2000-05-20 10:10:18} & yes & yes & \emph{no } \\
G29 & Ganymede & \texttt{2000-12-28 08:25:27} & yes & yes & \emph{no } \\
\rowcolor{lightgray}
C30 & Callisto & \texttt{2001-05-25 11:23:58} & yes & yes & \emph{yes} \\
I31 & Io       & \texttt{2001-08-06 04:59:21} & yes & yes & \emph{no } \\
I32 & Io       & \texttt{2001-10-16 01:23:21} & yes & yes & \emph{(?)} \\
I33 & Io       & \texttt{2002-01-17 14:08:23} & yes & yes & \emph{(?)}
\end{tabular}
    \caption{List of all targeted Galilean moon flybys with PWS/Electric-Survey data. 
    The data from all flybys (except E18) are available through the University of Iowa 
    \emph{das2} server end-point. The spacecraft ephemeris data is available for all 
    flybys except E15. The occultation assessment indicates if the occultation is 
    observed (\emph{yes} or \emph{no}) or unsure \emph{(?)}. The grey lines correspond 
    to the flybys shown in Figure \ref{fig:flyby_data} and described in \revised{detail} in 
    this study. Table adapted from Table 1 (Orbital Facts) of the 
    \texttt{CATALOG/GO\_MISSION.CAT} label file available from \citet{PWS_LPW_PDS}}
    \label{tab:gll-flyby-list}
\end{table*}

\subsection{Radio Emission \revised{modelling}}
\label{sec:ExPRES}
We model the location of the Jovian auroral radio sources visible at Galileo's location 
using ExPRES \citep[version 1.1.0,][]{louis_corentin_k_2020_4292002}). Our simulation runs 
are configured as follows: (a) we use the JRM09 magnetic field model \citep{Connerney:2018jx} 
together with the \citet{CAN_1981} current sheet model; (b) the sources are set every 
1$^\circ$ in longitude along active magnetic field lines of $\textrm{M-shell}=30$ (M-shell 
being the measure of the magnetic apex, i.e., the distance in Jovian radii ($R_J$), of the 
magnetic field line at the magnetic equator\revised{; it is different from the L-shell, for 
which the jovigraphic equator is considered for the magnetic apex}), corresponding to the 
main auroral oval \citep{Grodent:2015eo}; (c) the unstable electron temperature is set to 
5 keV \citep{Louarn:2017bc}; and (d) the location of the visible radio sources is 
\revised{modelled} with a temporal step of one minute. These parameters are fixed for all 
simulation runs used in this study. We also included Io-induced radio emissions in the 
simulation runs, with the same unstable electron distribution temperature (5 keV). The 
ExPRES configuration \revised{files are} available, as described in appendix 
\ref{app:supplementary}. 

When the observer is located near the magnetic equator, the radio source beaming pattern 
implies that the visible radio \revised{sources} are split into four cluster locations, called A, B, 
C and D, corresponding respectively to the North-Eastern, North-Western, South-Eastern and 
South-Western quadrant around Jupiter as seen from the observer \citep[see, e.g., Fig. 2 
of][for a definition]{2017A&A...604A..17M}. 

The simulation runs have been computed using the OPUS \citep[Observatoire de Paris UWS 
Server,][]{ 2021arXiv210108683S} instance operated by PADC\footnote{Paris Astronomical 
Data Centre:  \url{https://padc.obspm.fr} (Re3data record id: 
\url{http://doi.org/10.17616/R31NJMS9})} for the MASER (Measurement, Analysis and 
Simulation of Emissions in the Radio range) project \citep{Cecconi:2020bc}. OPUS is a 
framework running the Universal Worker Service (UWS) protocol \citep{2016ivoa.spec.1024H}. 
The ExPRES code is available for \revised{open access} run-on-demand from this 
interface\footnote{UWS MASER portal: \url{https://voparis-uws-maser.obspm.fr/client/}}. 

\subsection{Occultation \revised{modelling}}
\begin{figure}[ht]
    \centering\includegraphics[width=0.7\linewidth]{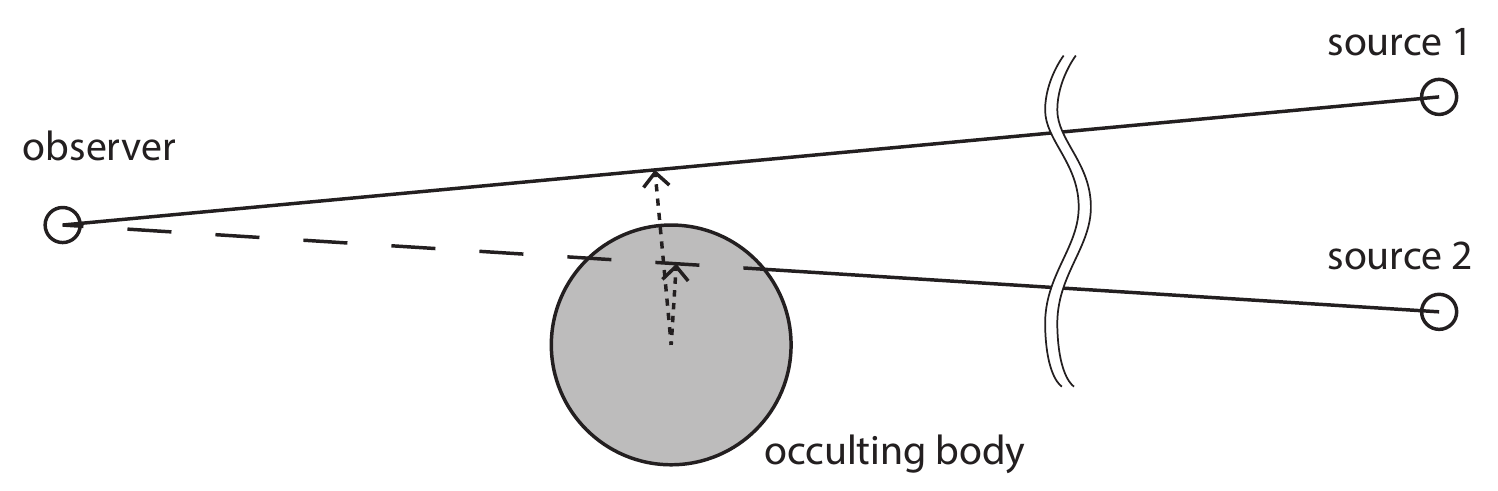}
    \caption{Simple geometric occultation scheme used in this study. The interception 
    distance (dotted segment) is computed as the distance between the center of the occulting 
    body and the line of sight between the observer and the source. In this case the source 1 
    is not occulted, while source 2 is.}\label{fig:occult}
\end{figure}

The occultation is computed using a simple geometric derivation of the intercept distance 
between the center of the Galilean moon and the straight lines passing by each visible 
modelled radio source and the observer. Any source with an intercept distance shorter than 
one moon radius is occulted, assuming a spherical moon, as sketched on Figure 
\ref{fig:occult}. 

\subsection{Occultation timing uncertainty}
\label{appendix:occultation-accuracy}

\begin{figure}
    \centering
    \includegraphics[width=0.5\linewidth]{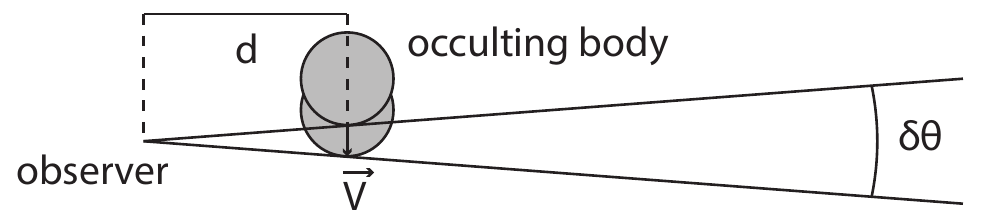}
    \caption{Simple flyby geometry used to evaluate the occultation timing uncertainty.
    The observer is noted ``S/C''. $\vec{V}$ is the relative velocity of the 
    observed moon with respect to the observer. $d$ is the distance of the observer
    to the moon. $\delta\theta$ is the angular uncertainty of the radio source location.}
    \label{fig:occult-accuracy}
\end{figure}

\revised{Using the radio source location accuracy estimated in 
\ref{sec:accuracy}, it is possible to estimate the uncertainty of the predicted 
occultation timings. Assuming the time uncertainty $\delta t$ to be sufficiently small, 
simplifying assumptions can be made: (i) the distance $d$ from the observer and the 
moon is constant; (ii) the relative velocity $\vec{V}$ between the observer and the 
moon is also constant, and is considered perpendicular to the line pointing to the 
radio source location. The angular uncertainty of the radio source location is 
$\delta\theta$. Figure \ref{fig:occult-accuracy} presents this simplified scheme. 
The observed moon is moving by a distance $\delta r = V \delta t$ to cover the 
angular uncertainty interval $\delta\theta$ as seen from the observer.  We can 
derive $\delta t$ as follows:
\begin{equation}
    \delta t \sim \frac{d\tan\delta\theta}{V}
    \label{eq:occultation-accuracy}
\end{equation}
With an angular uncertainty of the order of 1$^\circ$ (see \ref{sec:accuracy}), 
the previous equation evaluates as: $\delta t \sim 0.02 d / V$. This simplified 
estimation scheme corresponds to the worst case scenario. This measure of the 
uncertainty should thus be to taken as an upper limit.}

\section{Observations} 
All Galileo flybys have been analysed and modelled using ExPRES. In this section, we 
present the detailed modelling results corresponding to the highlighted rows of Table 
\ref{tab:gll-flyby-list}. Figures  \ref{fig:flyby_occultation_C30},  
\ref{fig:flyby_occultation_E12}, \ref{fig:flyby_occultation_G01} and 
\ref{fig:flyby_occultation_I24} present the results for these four flybys. The flybys 
are presented in order to show the simpler to the more complex cases. The GLL/PWS data 
(same data as in Figure \ref{fig:flyby_data}) are plotted together with the simulations 
of observable auroral radio emissions (described in Section \ref{sec:ExPRES}) separated 
into the four source types A, B, C and D (from white to black, respectively). The 
comparison of observations and modelled data shows that the simulations reproduce the 
Jovian radio occultation during the four flybys presented in Figure \ref{fig:flyby_data}.

\ref{app:supplementary} describes the supplementary material available for 
all flybys \citep{https://doi.org/10.25935/8zff-nx36}, which contains all the material 
used to conduct this study. For each Galileo flyby, we provide: (a) a figure showing the 
GLL/PWS data and the observable radio sources \revised{modelled} by ExPRES, and (b) 
movies showing a subset of Jovian radio sources (a selected sub-set of frequencies), 
as seen from Galileo (`pov' movies), or from the top of the Jovian system (`top' movies). 

\subsection{Callisto C30 flyby}

\begin{figure}[ht]
    \centering
    \includegraphics[width=0.8\linewidth]{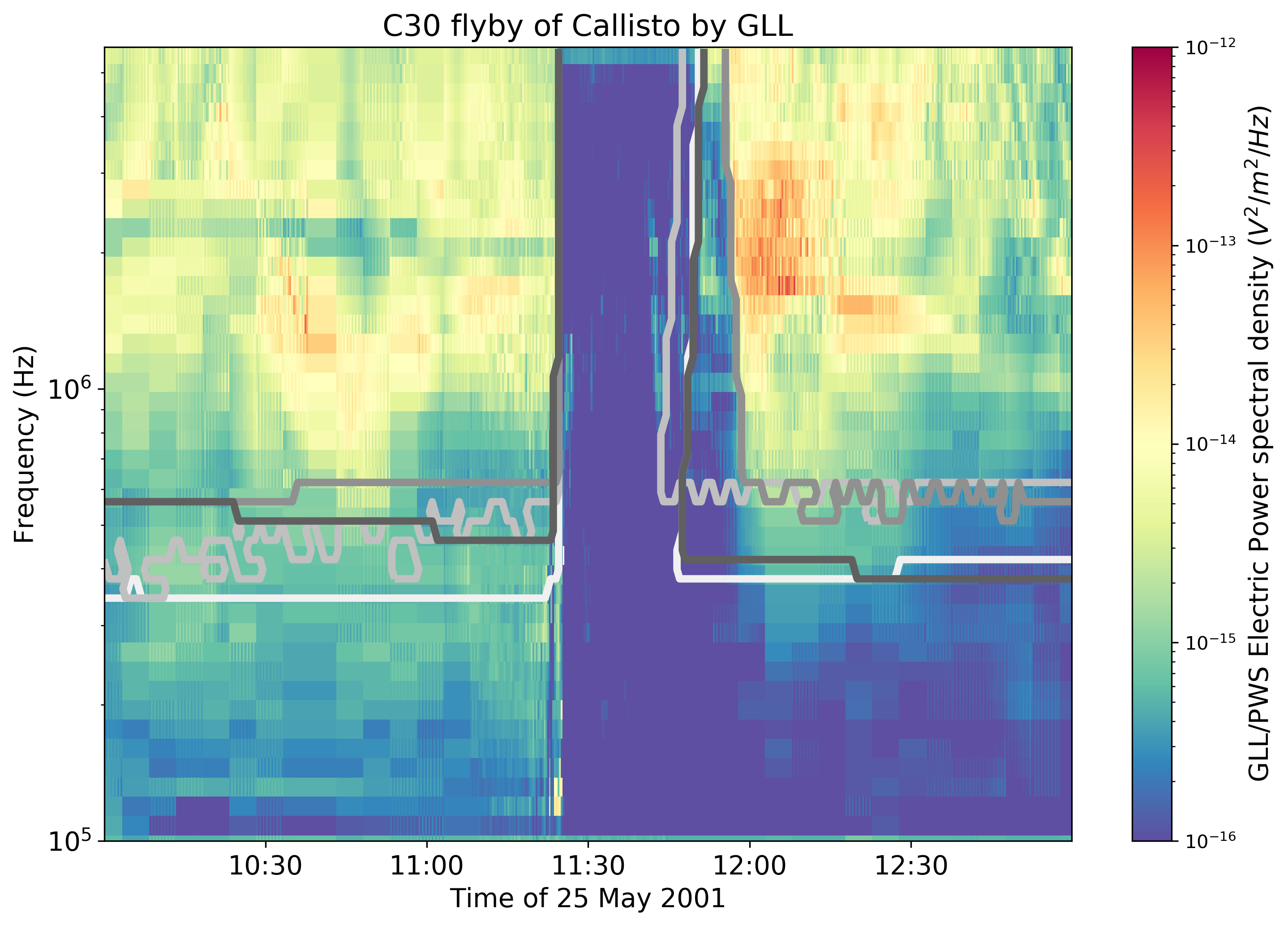}
    \caption{Superimposed GLL/PWS data and ExPRES simulations during Jovian radio 
    emission occultations by Callisto (flyby C30). The four types of emission (A, B, 
    C, D) are separated (from white to dark grey, resp.)}
    \label{fig:flyby_occultation_C30}
\end{figure}

The occultation occurring during the C30 flyby of Callisto is displayed in Figure 
\ref{fig:flyby_occultation_C30}. The ingress occultation time is very well reproduced 
(within one minute accuracy). All sources are occulted simultaneously, with radio 
emissions intensity dropping instantly, at $\sim$11:25 SCET. At egress, we observe 
first a faint rise of the emission intensity from $\sim$11:50 to $\sim$12:00, and 
then a sudden return to maximum intensity at $\sim$12:00. This egress phase can be 
visualised in the supplementary material available for this flyby: 
\url{https://doi.org/10.25935/8ZFF-NX36#C30}. The frames between 11:44 to 11:59 of 
the `pov' movie clearly shows the various sources reappearing one after the other: 
first, the B sources, then the A and D sources simultaneously and finally the C 
sources. The predicted reappearance of C sources perfectly coincides with the observed 
full egress phase. This leads to two observations: (i) the main radio contribution is 
that of the C sources at the time of observation, and (ii) the radio sources are 
occulted by the moon's surface (or very close to it).

The low-frequency cut-off is not perfectly simulated (especially on the ingress side), 
with a \revised{shift} of about $200$~kHz \revised{with respect to the observations. The 
error of a few 10s-100s kHz is probably due to an incorrect estimation of the electron 
density around Jupiter. In addition to this spectral shift, some components show an 
erratic cut-off line shape (B sources --light grey-- throughout the interval, except 
during the occultation, and C sources --medium grey-- after the occultation). It is not 
fully understood why and under which conditions, some components show a rather stable 
cut-off line, and some other a more erratic one. However, it is noticeable that the 
erratic features of the B sources cut-off line look like the lower end of vertex-early
arcs (e.g., at $\sim$10:10, $\sim$10:25 or $\sim$11:45), and those of the C sources 
look like those of vertex-late arcs (e.g., at $\sim$12:15).}  

An attenuation feature is observed starting at $\sim$2.5~MHz, and 
decreasing to $\sim$2~MHz at ingress. This corresponds to the attenuation lanes 
described in the introduction section. In addition, during the full occultation 
period of time, unexpected faint and sporadic signals are observed between 
$\sim$800~kHz and $\sim$2~MHz.

\subsection{Europa E12 flyby}

\begin{figure}[ht]
    \centering
    \includegraphics[width=0.8\linewidth]{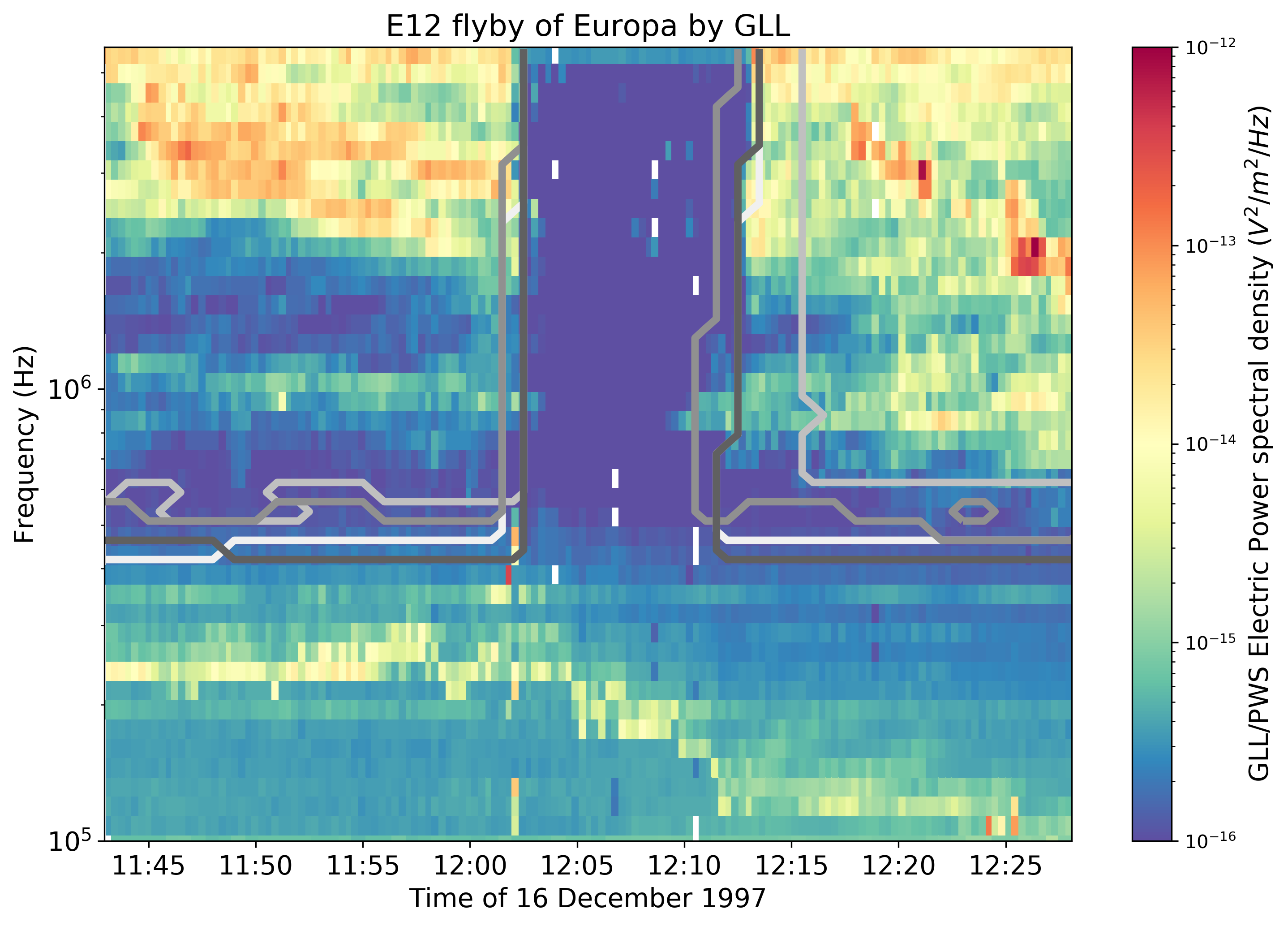}
    \caption{Superimposed GLL/PWS data and ExPRES simulations during Jovian radio 
    emission occultations by Europa (flyby E12). The four types of emission (A, B, C, 
    D) are separated (from white to dark grey, resp.)}
    \label{fig:flyby_occultation_E12}
\end{figure}

The E12 \revised{Europa} flyby (Fig. \ref{fig:flyby_occultation_E12}) is similar to C30 
Callisto's flyby case, where all sources are occulted simultaneously at ingress. 
Faint and sporadic radio signatures are observed during the full occultation phase 
(as in the case of C30). At egress, the simulation well reproduces the end of the 
occultation, except in the $\simeq$ [$800$-$1000$]~kHz frequency range where we 
observed emission during the modelled occultation. \revised{The full E12 flyby data is 
available at:} \url{https://doi.org/10.25935/8zff-nx36#E12}

The attenuation lane feature is observed at and below $\sim$2~MHz before ingress, 
with a corresponding feature after egress, up to $\sim$12:20. At about $\sim$1~MHz, 
an intensification is also observed, before ingress and after egress (corresponding 
to the aforementioned prediction mismatch), and is probably linked to the attenuation 
lanes.

The spectral line observed between $\sim$250~kHz (at the beginning of the interval) 
and $\sim$100~kHz (at the end) corresponds to the local plasma upper-hybrid frequency 
line \revised{($f_\mathit{UH}=(f_{pe}^2 + f_{ce}^2)^{1/2}$, with $f_{pe}$ and $f_{ce}$, 
the plasma frequency and the electron cyclotron frequency, respectively).}


\subsection{Ganymede G01 flyby}

\begin{figure}[ht]
    \centering
    \includegraphics[width=0.8\linewidth]{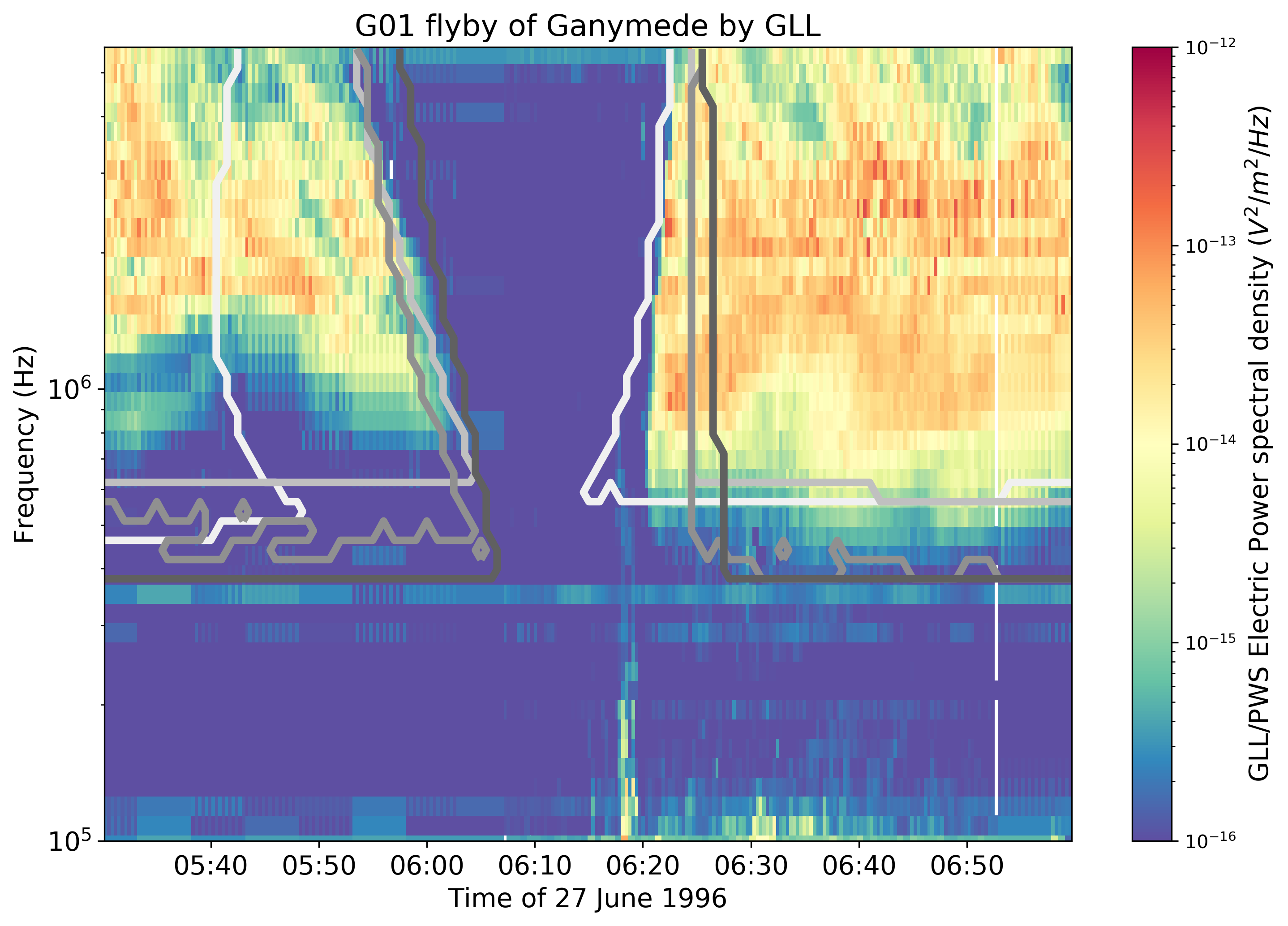}
    \caption{Superimposed Galileo  PWS data and ExPRES simulations during Jovian 
    radio emission occultations by Ganymede (flyby G01). The four types of emission 
    (A, B, C, D) are separated (from white to dark grey, resp.)}
    \label{fig:flyby_occultation_G01}
\end{figure}

\begin{figure*}[th]
    \centering
    \includegraphics[width=\linewidth]{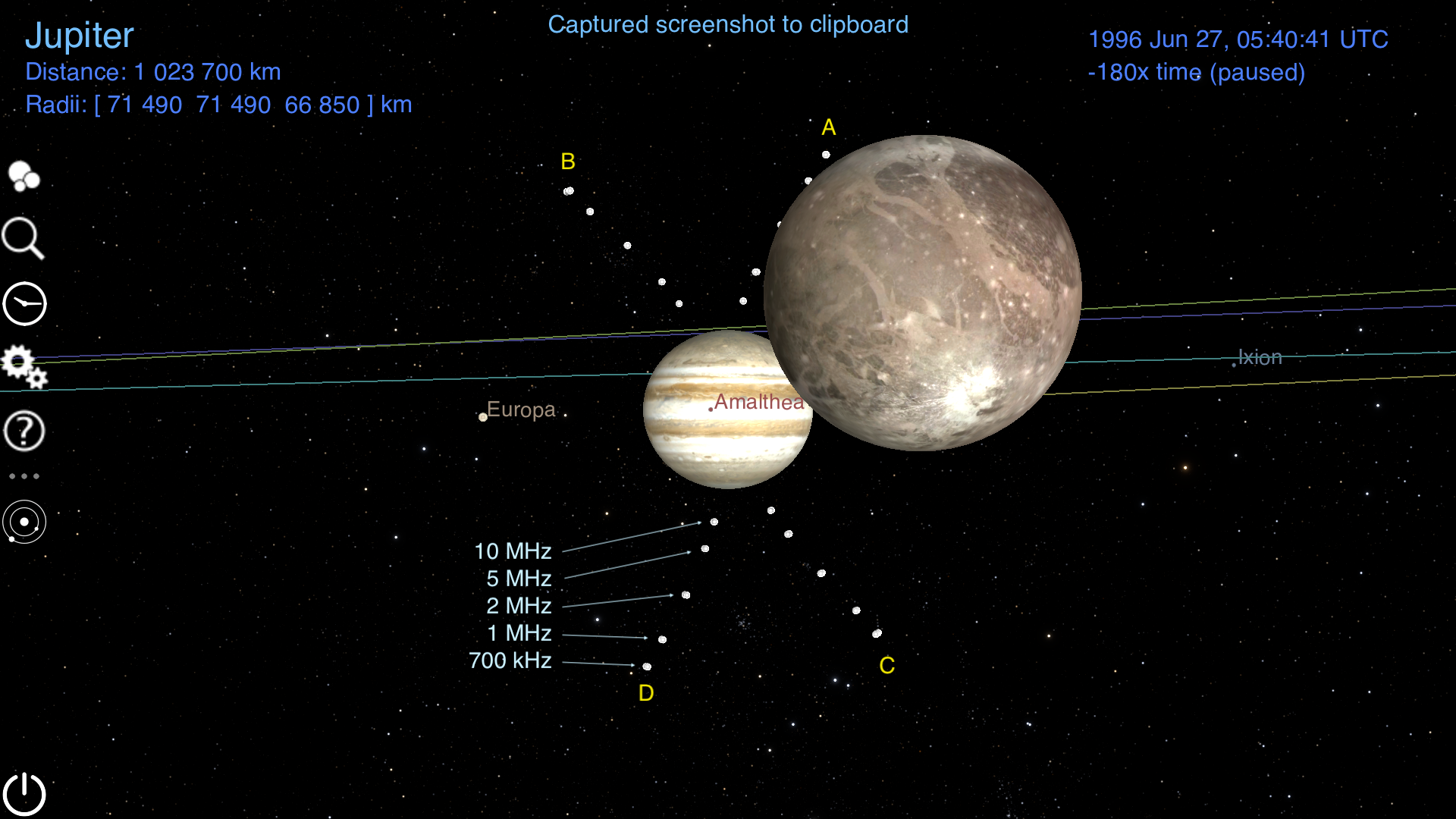}
    \caption{G01 Flyby visualised in the Cosmographia tool. The scene is set with an 
    observer on the Galileo spacecraft, pointing to Jupiter. Ganymede is in the field of 
    view. The ExPRES-modelled visible radio sources are also shown, at 700 kHz, 1 MHz, 2 
    MHz, 5 MHz and 10 MHz. The radio sources are naturally grouped in four sets (named A, 
    B, C and D). A the time of the snapshot (\texttt{1996-06-27T05:40:41} SCET), \revised{the 
    ingress phase is starting, with the occultation of radio sources} at 2 MHz in the A-group.}
    \label{fig:cosmo-g01}
\end{figure*}

Figure \ref{fig:flyby_occultation_G01} displays the occultation modelling of the Jovian 
radio emissions during the G01 flyby of Ganymede. At ingress, unlike the previous two 
cases, all sources are not occulted simultaneously. First the A sources are occulted, 
then the B, C and D sources, with the beginning of the full occultation.
Before observed egress, a broadband noise burst at $\sim$06:18, going up to $\sim$800 
kHz. This has been interpreted as the signature of Ganymede's magnetopause crossing by 
\citet{Gurnett:1996tt}. At egress, the modelled reappearance of the A sources (white 
line) well \revised{reproduces} the end of the observed occultation at frequencies 
higher than a few MHz. At lower frequencies, the egress occurs later than predicted. 
At $\sim 700$ kHz, it is observed at $\sim$06:21, and predicted at $\sim$06:16. 
\revised{This mismatch is likely due to refraction of radio waves in Ganymede's 
atmosphere or ionosphere.}

During this flyby, the radio sources are not occulted simultaneously at both egress 
and ingress, and not at the same time on the whole spectral range, which is due to 
the trajectory of Galileo. Figure \ref{fig:cosmo-g01} shows a modelled image, from 
the Galileo spacecraft point of view, at the beginning of the G01 occultation sequence. 
This frame is extracted from the G01 movie available in supplementary material at: 
\url{https://doi.org/10.25935/8zff-nx36#G01}.

\subsection{Io I24 flyby}

\begin{figure}[ht]
    \centering
    \includegraphics[width=0.8\linewidth]{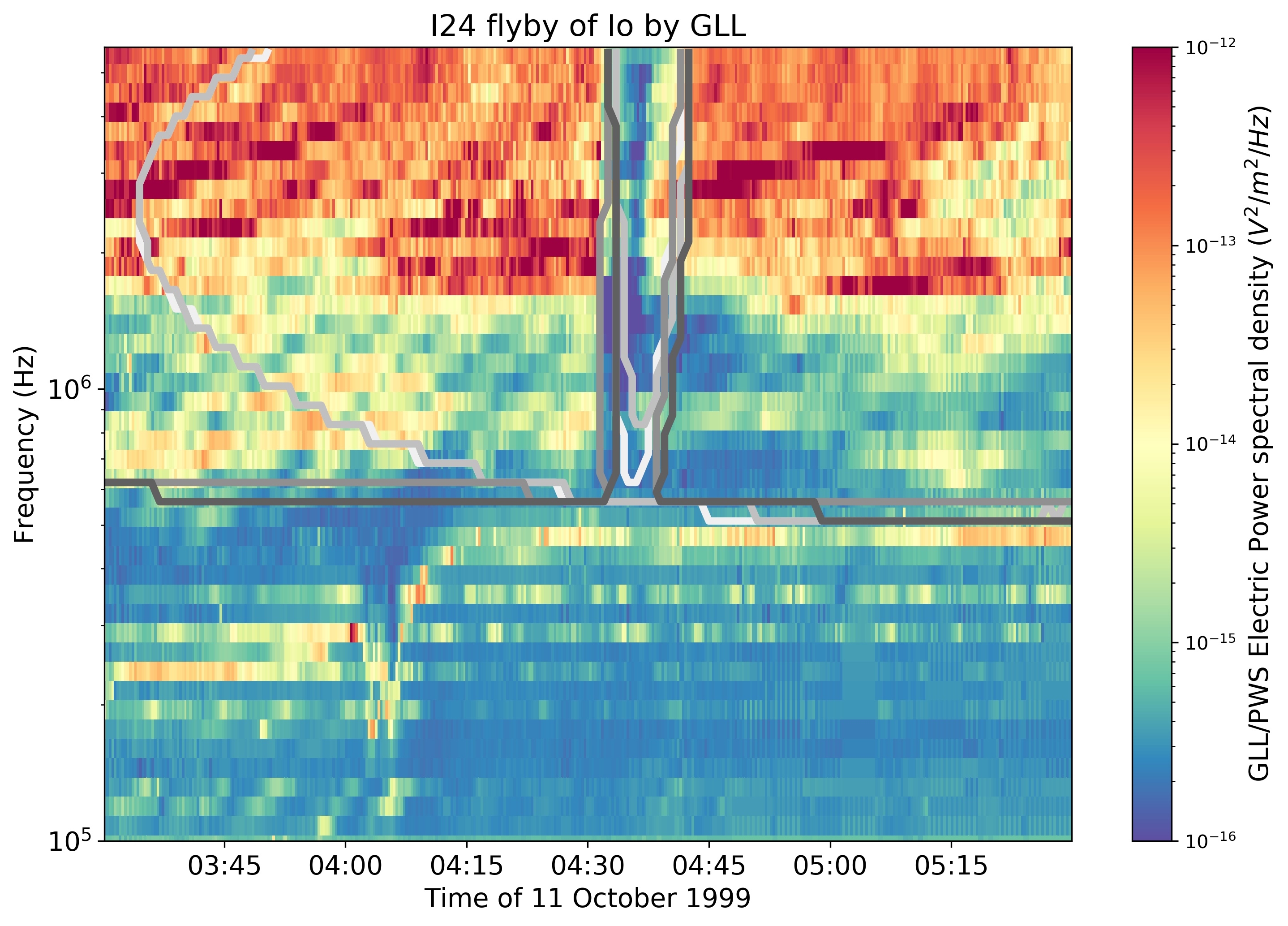}
    \caption{Superimposed GLL/PWS data and ExPRES simulations during Jovian radio 
    emission occultations by Io (flyby I24). The four types of emission (A, B, C, D) 
    are separated (from white to dark grey, resp.)}
    \label{fig:flyby_occultation_I24}
\end{figure}

Figure \ref{fig:flyby_occultation_I24} displays the occultation during the I24 Io's 
flyby. At ingress, the occultation of the higher intensity is well modelled. We 
observe radio signals during the modelled full occultation, and the observed egress 
seems to occur earlier than the prediction. Intense radio arcs are visible above 2~MHz, 
showing the lower frequency part of vertex-late arcs. The radio signal is attenuated 
below 1 to 2~MHz, especially after the flyby. The $f_{UH}$ line is also observed 
($\sim$300~kHz to $\sim$500~kHz). \revised{The full I24 flyby data is 
available at:} \url{https://doi.org/10.25935/8zff-nx36#I24}



This Io flyby also shows a noticeable feature. The modelled Northern radio sources 
(A and B, respectively in white and light-grey) are not observed at the beginning of 
the studied interval. This is due to the ``Equatorial Shadow Zone'' (ESZ) effect 
reported at Saturn \citep{lamy_JGR_08b}, see \ref{appendix:esz} for more 
details. 

\subsection{Other flybys}

\begin{figure}[ht]
    \centering
    \includegraphics[width=0.8\linewidth]{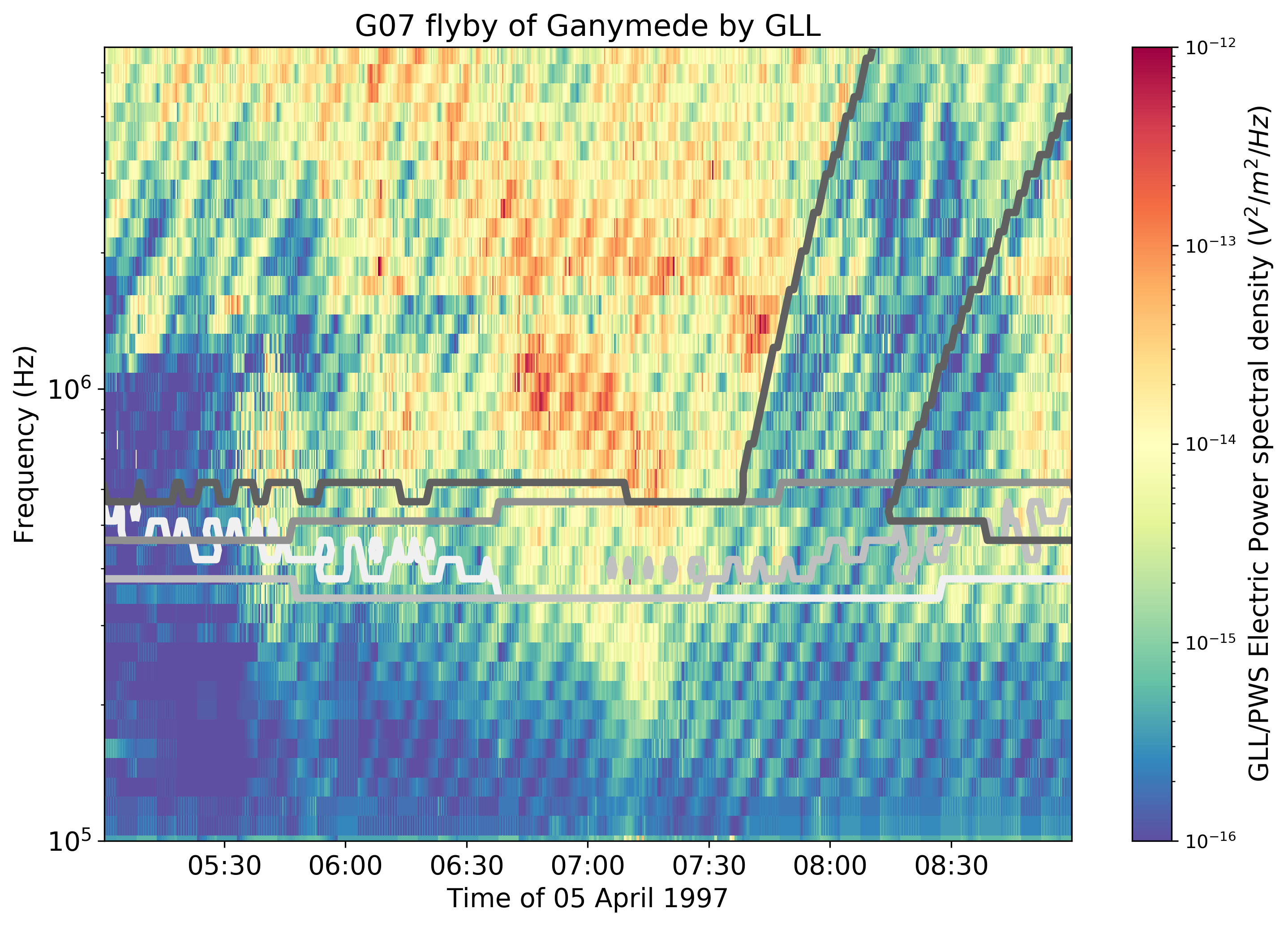}
    \caption{Superimposed Galileo  PWS data and ExPRES simulations during Jovian 
    radio emission occultations by Ganymede (flyby G07). The four types of emission 
    (A, B, C, D) are separated (from white to dark grey, resp.)}
    \label{fig:flyby_occultation_G07}
\end{figure}

In the case of flybys with partial occultations (e.g., G02, G07, E11, E26 or G28), 
the Jovian radio signals are observed during the flyby, with no full occultation 
interval. During the G07 flyby, only D sources are occulted. The GLL/PWS data (see 
\url{https://doi.org/10.25935/8zff-nx36#G07}) shows an attenuation of the Jovian 
radio signals that fits the predicted occultation (see Figure 
\ref{fig:flyby_occultation_G07}), hence suggesting that several radio sources were 
active, including the D sources. 

Most of the other flybys (e.g., C03, E04, E06, G08, C10, E19) are occurring on the 
Jovian-facing side of the moon, where no occultation can occur.

\section{Results and Discussion} 
\label{sec:Discussion}

\subsection{Temporal Occurrence and Spectral Coverage of Jovian Auroral Radio Signals}

The first result of this study is the fact that Jovian auroral radio emissions are 
present almost continuously, although this may not be obvious in some observational 
data sets, due to the limited sensitivity of instruments. Ground based radio 
telescopes (such as the Nan\c cay \revised{Decametre} Array, in France) provide the 
longest observation collections \citep{2017A&A...604A..17M}, but the distance to 
Jupiter limits the detection of low intensity events despite their intrinsic 
sensitivity. Space borne radio instruments (such as Voyager/PRA, Cassini/RPWS, 
GLL/PWS or Juno/Waves), provide observations from a closer range to Jupiter. Cassini 
and Galileo data show quasi-continuous radio signals (as shown on Figure 
\ref{fig:cas-gll-continuous-radio}). The ExPRES \revised{modelling} of auroral radio 
sources, configured with radio sources every 1$^\circ$ in longitude, is consistent 
with this observation (with the noticeable exception of the ESZ, for an observer 
located around Io's orbital distance to Jupiter). 

The occultation \revised{modelling} analysis shows that all sources must be occulted 
\revised{at the same time} in order to remove the natural radio signature of Jupiter's 
magnetospheric activity. It is also noticeable that faint emissions are still visible 
during some occultations (e.g., during the G01, E12 and C30 Galileo flybys). 

The JUICE/RPWI and RIME instruments will observe \revised{in the same radio environment}. 
The JUICE/RIME and Cassini/RPWS instruments have similar antenna characteristics, leading 
to an antenna resonance at about $\sim$9 MHz \citep{zarka_JGR_04, Bruzzone:2013ge}. 
It is thus very likely that JUICE/RIME will observe similar signals as those shown 
on Figure \ref{fig:cas-gll-continuous-radio} (with a spectral range restricted around 
9 MHz).

The low frequency limit is usually well predicted by ExPRES, as shown on Figures 
\ref{fig:flyby_occultation_C30}, \ref{fig:flyby_occultation_E12}, 
\ref{fig:flyby_occultation_G01}, \ref{fig:flyby_occultation_I24} and 
\ref{fig:flyby_occultation_G07}. The ExPRES low frequency emission limit is determined 
by the CMI theory: the emission can occur only when $f_{pe}/f_{ce} < 0.01$. This 
finding implies that our \revised{modelling} of the magnetic field and plasma density is 
consistent with the radio observations. Discrepancies (e.g., during flyby C30) may be 
related to the \revised{modelling} of the two aforementioned characteristic frequencies: 
$f_{pe}$ depends on the magnetospheric current sheet model; and $f_{ce}$ is determined by 
the magnetic field model. The JRM09 magnetic field model is derived from the Juno magnetic 
measurements in the polar regions of Jupiter. This model is thus perfectly adapted 
for our application. Conversely, the current sheet model could be improved. The results 
of this study will be \revised{checked} and updated \revised{if required when the new 
\citet{Connerney:2020fv} model is included into ExPRES.} The effect of the Solar Wind 
conditions may also play a role \citep[as studied by][]{Hess:2012dx} \revised{at the 
lowest frequencies.}

\subsection{Radio source location}
\begin{figure}[ht]
    \centering
    \includegraphics[width=0.8\linewidth]{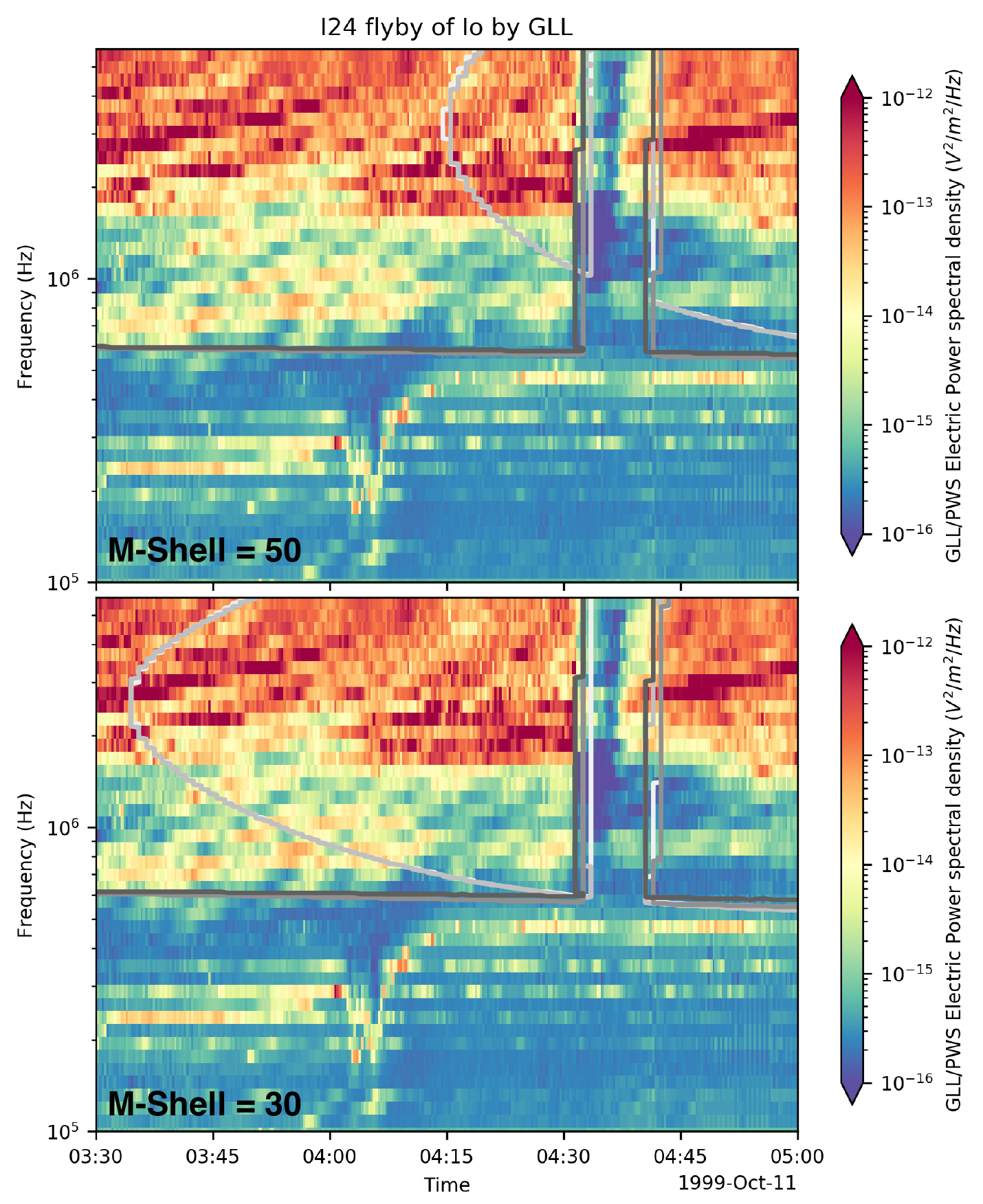}
    \caption{Comparison between the Jovian radio occultation during the I24 flyby for 
    sources on magnetic field lines at M shell M=50 (top panel) and M=30 (bottom panel).}
    \label{fig:compa_mshell_50_30}
\end{figure}

Figure \ref{fig:compa_mshell_50_30} displays the occultation of the Jovian radio 
sources during the flyby I24, for two sets of simulation runs. In the top and bottom 
panels, the sources are located on the magnetic field lines with an M-shell of 50 and 
30 $R_J$, respectively, with the larger M-shell, corresponding to radio sources located 
at higher magnetic latitudes. We observe no difference for the occultation prediction. 
\revised{This finding implies that updating the current sheet model \citep{Connerney:2020fv} 
should have a limited impact on the results of this study. }

The difference in the low cut-off frequency of the simulated emissions is very small 
(with a slightly lower low cut-off frequency for sources with M-shells of 50 $R_J$). 
The main difference is observed for the A and B sources (white and light-grey curve), 
with the ESZ feature occurring at a different time. 

\revised{An assessment of the radio source location prediction accuracy is presented in 
\ref{sec:accuracy}. We show that the overall angular accuracy of the radio source 
location is below 1$^\circ$ at 10 MHz for observations at Europa, Ganymede and Callisto. 
The 1$^\circ$ accuracy threshold is met at 5 MHz at Europa, 2 MHz at Ganymede, and 700 
kHz at Callisto. See Table \ref{tab:radio-source-location-accuracy} for detailed 
results.}

\subsection{Occultation \revised{modelling} accuracy and propagation effects}

The ExPRES modelling \revised{accurately predicts} the observed occultation 
(e.g., C30 flyby, or E12 and I24 flyby ingress). The ExPRES simulation runs 
have been configured with a 1 minute sampling step, which sets the general 
temporal sampling of this study. \revised{Table \ref{tab:gll-accuracy} shows 
the modelled occultation timing accuracy, using Equation \ref{eq:occultation-accuracy}. 
All values are below one minute, justifying the one minute sampling precision used
in this study.} The discrepancies between the predictions and
observation have to be further studied. The mismatch mostly occurs at frequencies 
lower than $\sim$1~MHz. In this range, propagation effects (such as refraction 
effects) are known to occur, with, e.g., attenuation lanes. We observe them in 
all studied flybys, e.g., on E12, where the Jovian auroral radio waves are attenuated 
below $\sim$2~MHz, with an intensification at $\sim$1~MHz.   

\begin{table}[ht]
    \begin{tabular}{l|l|l|rrrrr}
        Flyby & Phase & Date Time (SCET) &  \multicolumn{5}{c}{Temporal accuracy (s)} \\
        \hline
        G01 & ingress & 1996-06-27 06:00:00 & 55 & 45 & 32 & 19 & 13 \\
        G01 & egress & 1996-06-27 06:20:00 & 22 & 18 & 13 & 7 & 5 \\
        \hline
        E12 & ingress & 1997-12-16 12:02:00 & 14 & 12 & 8 & 5 & 4 \\
        E12 & egress & 1997-12-16 12:12:00 & 29 & 24 & 17 & 11 & 8 \\
        \hline
        I24 & ingress & 1999-10-11 04:33:00 & 23 & 19 & 14 & 9 & 6 \\
        I24 & egress & 1999-10-11 04:42:00 & 49 & 40 & 28 & 18 & 12 \\
        \hline
        C30 & ingress & 2001-05-25 11:25:00 & 5 & 4 & 3 & 1 & 1 \\
        C30 & egress & 2001-05-25 11:50:00 & 26 & 20 & 15 & 8 & 5 \\
        \hline
        \hline
        \multicolumn{3}{r|}{Frequency (MHz)} &  0.7 & 1.0 & 2.0 & 5.0 & 10.0
    \end{tabular}
    \caption{Estimation of the temporal uncertainty of the predicted occultation 
    timing in seconds.}
    \label{tab:gll-accuracy}
\end{table}

\begin{figure*}[ht]
    \centering
    \includegraphics[width=\linewidth]{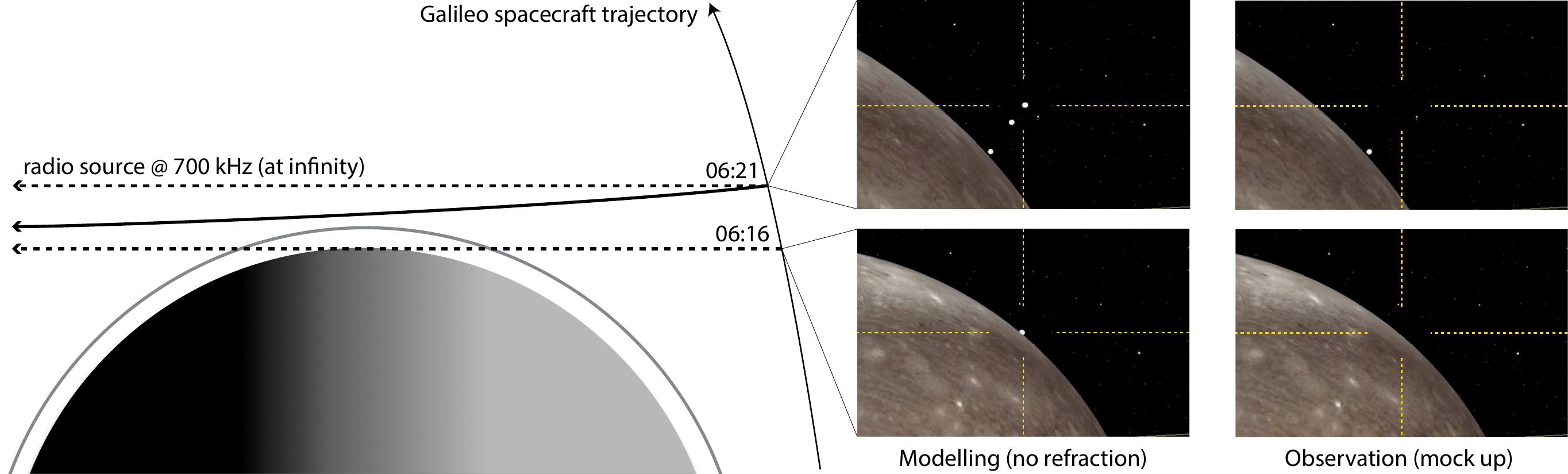}
    \caption{Sketch of G01 configuration, with a modelled radio source at 700 kHz, 
    assumed to be at infinity on the left-hand side. The plain line is a sketched 
    refracted ray path. The plain grey line represents the ionosphere of the moon. 
    The dashed lines are the straight line propagation (no refraction) for the same 
    source. Two locations of the Galileo spacecraft trajectory are illustrated on the 
    right-hand side of the Figure. The ``Modelling (no refraction)'' column shows 
    frames extracted from the G01 `pov' movie at 06:16 and 06:21. The right-most column 
    (``Observation'') shows \revised{a mock-up of the radio source observations, including 
    refraction effects} at low frequencies.}
    \label{fig:G01-refraction}
\end{figure*}

In the case of G01, the egress is observed several minutes after the predicted egress 
time. \revised{Since refraction in the moon's atmosphere and ionosphere is bending 
away optical paths \citep[see, e.g., ][]{Colin:1972vg}, the apparition of the source at 
egress is observed later than compared to the straight line propagation case.} 
Figure \ref{fig:G01-refraction} shows how the same observations can be interpreted 
considering refraction effects on the ionosphere of the moon.

In Figures \ref{fig:flyby_occultation_C30}, \ref{fig:flyby_occultation_E12} and 
\ref{fig:flyby_occultation_G01}, we observe faint and sometimes sporadic radio signals, 
which are visible during the occultation interval, despite the predicted full occultation. 
Since the moon is geometrically occulting all the Jovian radio sources, refraction 
effects must be taken into account to interpret the observation. These effects can occur 
either in the moon's atmosphere and ionosphere, in the Io plasma torus or in the 
magnetospheric plasma sheet. ExPRES assumes a straight line propagation between the 
radio source and the observer. 

The two latter results (occultation ingress and egress prediction mismatch and faint 
signals during full occultation) indicate that propagation effects play an important 
role in the fine understanding of the Galilean radio occultations. Further analysis 
requires the coupling to a ray-tracing code, such as ARTEMIS-P \citep[\emph{Anisotropic 
Ray Tracer for Electromagnetism in Magnetosphere, Ionosphere and Solar wind including 
Polarization}, ][]{Gautier:2013uh}.

\revised{\ref{appendix:refraction} is presenting a preliminary study estimating the range of 
parameters (e.g., altitudes above the surface, or plasma densities), in which the 
refraction effects are likely to occur. The estimation is based in the currently 
available Galilean moon's ionosphere models (see Table \ref{tab:iono_models}). 
Using those environment models, the refractive index is computed, using the 
Appleton-Hartree equation (see Equation \ref{eq:appleton-hartree}). From Figure 
\ref{fig:n2-moons} and Equation \ref{eq:appleton-hartree-y-0}, we can derive rough 
estimates of the accessible parameter space with observations at each of the observation 
frequencies used in this study, as shown in Table \ref{tab:parameter-space}. A first
conclusion of this estimation is that no significant refraction is expected at 10 MHz at 
Europa, Ganymede and Callisto, so that the simple occultation modelling presented in 
this study should be applicable around this frequency.}

\section{Usage for the JUICE mission planning tools} 

The science planning activity, coordinated by the JUICE Science Operations Center (JUICE 
SOC), relies on the identification, at each point in time, of the science \revised{observation}
opportunities, using science models or/and geometry. For JUICE, some of those 
\revised{opportunities} depend on the Jupiter Radio emission simulation. Ionosphere
\revised{characterisation,} active and/or passive radar sounding activities opportunities 
can benefit from an accurate simulation of Jupiter radio emission.

Since the ExPRES \revised{code} can provide as \revised{by-products} the \revised{locations} 
of the radio sources as seen from the spacecraft for a range of frequencies (i.e., 
1--40 MHz), the \revised{JUICE SOC} has implemented a standalone tool, which wraps the 
ExPRES \revised{code} and \revised{identifies} science opportunity windows based on the 
radio source position.  This can be used to identify opportunity windows for one of the 
high priority science objectives of the \revised{JUICE} RPWI instrument, the icy moon 
ionosphere \revised{characterisation} through ionosphere refraction/distortion measurement.

The opportunity periods currently generated to drive the corresponding RPWI measurements 
are linked to the ingress and egress occultation events of the Jupiter \revised{radio sources}, 
where at least one Jupiter radio sources \revised{reaches JUICE} with a frequency between 
0.1--5 MHz crossing the moon ionosphere (with \revised{a thickness of $\sim$100~km}).

\begin{figure*}[ht]
    \centering
    \includegraphics[width=\linewidth]{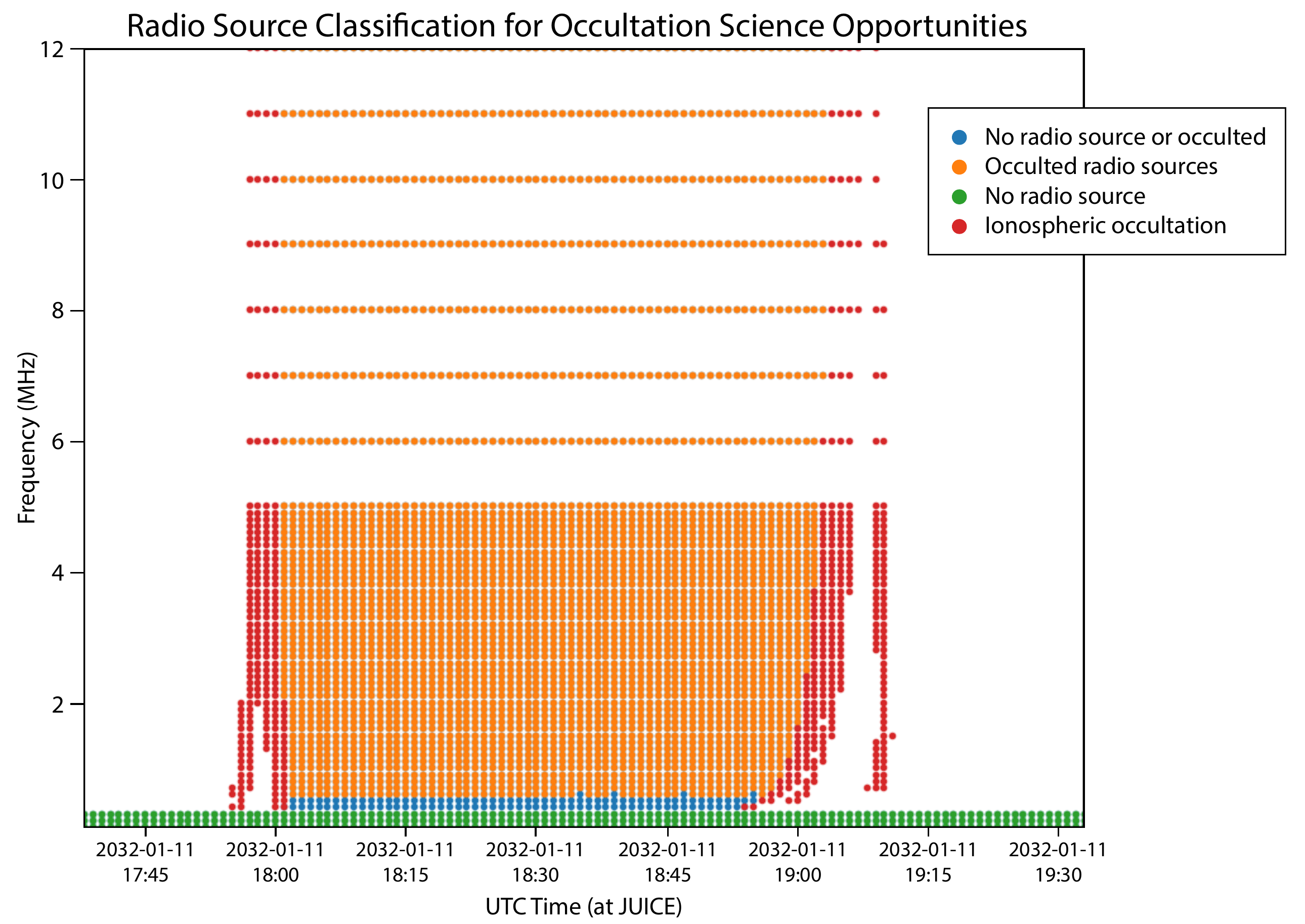}
    \caption{\revised{Classification of observation opportunities during the the last Callisto flyby 
    (21C13) of orbital scenario CReMA 3.0. The color coded dots represent the Jupiter auroral 
    radio source occultation state \revised{as} a function of time and frequency. Red dots means 
    that at least one of the source type (A, B, C and D) is visible from JUICE and that 
    the line of sight between the source and the spacecraft goes through the ionosphere 
    within 0-100 km altitude above Callisto surface. The blue, orange and green dots mean 
    that no radio source is visible by JUICE at those frequency ranges. Here ``No Radio 
    Source'' means that no signal from Jupiter radio sources should be observed: either because
    JUICE is outside the beam of all modelled radio sources, or because the model predicts that the
    radio source can't produce radio waves (e.g., in the lower frequency range)}.}
    \label{JUICE SOC-fig1}
\end{figure*}

In \revised{figure} \ref{JUICE SOC-fig1}, the Jupiter auroral radio source as seen from JUICE 
during 21C13 (the last Callisto flyby of the tour) as a function of frequency (MHz)
and UTC time is displayed: the red dots correspond to opportunity for ionosphere 
\revised{characterisation}, i.e., whenever one of the radio source types (A, B, C or D) is seen 
by JUICE with a line of sight passing through the ionosphere within 0--100\,km. This 
section is using the CReMA 3.0 trajectory scenario.

\begin{figure*}[ht]
    \centering
    \includegraphics[width=\linewidth]{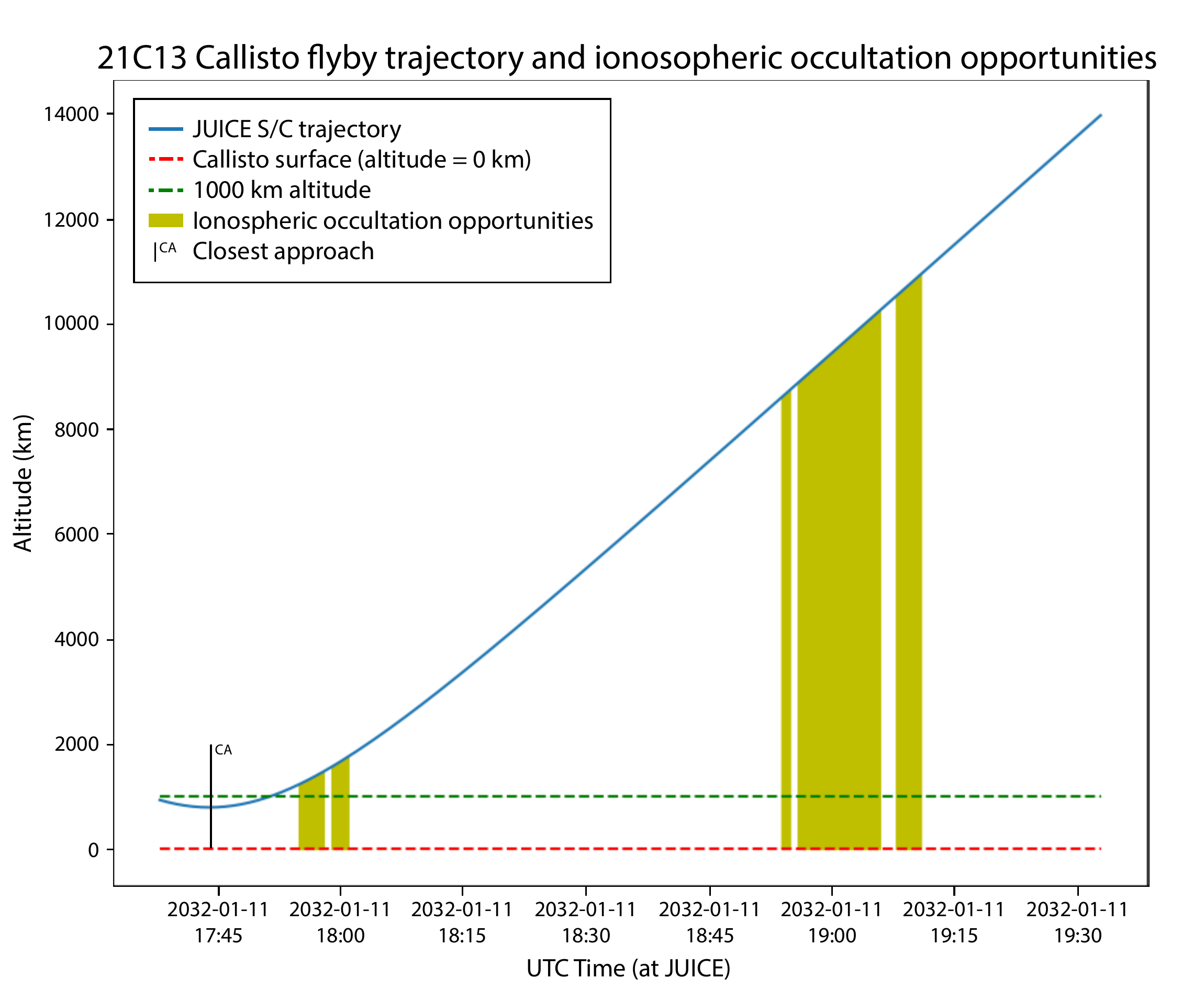}
    \caption{\revised{JUICE} spacecraft \revised{altitude above the surface of Callisto} in km for 
    the 21C13 Flyby (blue line); \revised{Callisto's} surface (resp. the 1000km altitude above surface) is 
    represented by the red dashed line (resp.\ green dashed line), and the ionosphere 
    \revised{characterisation} opportunities windows are filled in \revised{green} background. \revised{The
    figure only shows the outbound part of the 21C13 flyby, since the radio occultation opportunities are 
    only present during this phase.}}
    \label{JUICE SOC-fig2}
\end{figure*}

Figure \ref{JUICE SOC-fig2} shows the resulting \revised{ionospheric characterisation} opportunities 
as a function of the \revised{JUICE} spacecraft altitude in km (green background). There are a few 
gaps within the ingress and egress \revised{ionospheric characterisation} opportunities. Those gaps of 
1-2 minutes are ignored to compute the final \texttt{iono\_ingress} and \texttt{iono\_egress} 
envelops used for the \revised{observation} planning. Table \ref{JUICE SOC-tab1} lists the RPWI 
in-situ and radio measurement sequence for the 21C13 scenario based on the Closest Approach 
(CA), and the ingress and egress windows envelope for ionosphere \revised{characterisation}. 
\revised{Using equation \ref{eq:occultation-accuracy}, we can estimate the uncertainty of the 
occultation timing (see Table \ref{tab:21C13-uncertainty})}.

\begin{table*}[ht]
    \centering
    \begin{tabular}{l|c|c|c|c|c}
              & Date Time & Frequency & Altitude & Velocity & Uncertainty\\
        Event & (SCET)    & (MHz)     & (km)     & (km/s)   & (s)\\
        \hline
        Full Occ.\ Start & 2032-01-11 18:00:22 & 10.0 & 1686.7 & 2.88 & 5 \\
        Full Occ.\ End & 2032-01-11 19:03:22 & 10.0 &	9880.1 & 2.45 & 18 \\
    \end{tabular}
    \caption{\revised{Full occultation events at 10 MHz for JUICE flyby 21C13 of Callisto, with spacecraft altitude, relative velocity, and occultation timing uncertainty.}}
    \label{tab:21C13-uncertainty}
\end{table*}

\begin{table*}[ht]
    \begin{tabular}{l|l|r|l|l}
UTC Date \& Time            & Reference Event & Relative Time  & In-situ& Radio\\ \hline
\texttt{2032-01-11T05:44:05}&\texttt{CA}&\texttt{$-$12:00:00}& slow	  &full\\
\texttt{2032-01-11T08:14:05}&\texttt{CA}&\texttt{$-$09:30:00}& normal &full\\
\texttt{2032-01-11T17:34:05}&\texttt{CA}&\texttt{$-$00:10:00}& burst  &full\\
\texttt{2032-01-11T17:54:05}&\texttt{CA}&\texttt{$+$00:10:00}  & normal &full\\
\texttt{2032-01-11T17:55:00}&\texttt{iono\_ingress}&\texttt{$+$00:00:00}&normal & burst\\
\texttt{2032-01-11T18:01:00}&\texttt{iono\_egress} &\texttt{$+$00:00:00}&normal & full\\
\texttt{2032-01-11T18:54:00}&\texttt{iono\_ingress}&\texttt{$+$00:00:00}&normal & burst\\
\texttt{2032-01-11T19:11:00}&\texttt{iono\_egress} &\texttt{$+$00:00:00}&normal & full\\
\texttt{2032-01-12T03:14:05}&\texttt{CA}&\texttt{$+$09:30:00}  & slow   &full\\
\texttt{2032-01-12T05:44:05}&\texttt{CA}&\texttt{$-$12:00:00}& slow   &full
    \end{tabular}
    \caption{RPWI in-situ and radio observations mode sequence during the 21C13 Callisto 
    flyby scenario. Some mode changes are scheduled w.r.t.\ Closest Approach event (CA), 
    while ionosphere \revised{characterisation} observations are scheduled w.r.t.\ ingress and egress 
    events as identified using Express. \revised{The last two column indicates which operating mode is 
    for the \emph{In-situ} or \emph{Radio} sub-systems of RPWI. Each mode has a specific data rate, as 
    well as a specific power consumption. Without explaining the details of each modes, this table shows 
    how the various modes are activated sequentially.}}
    \label{JUICE SOC-tab1}
\end{table*}

\begin{figure*}[ht]
    \centering
    \includegraphics[width=\linewidth]{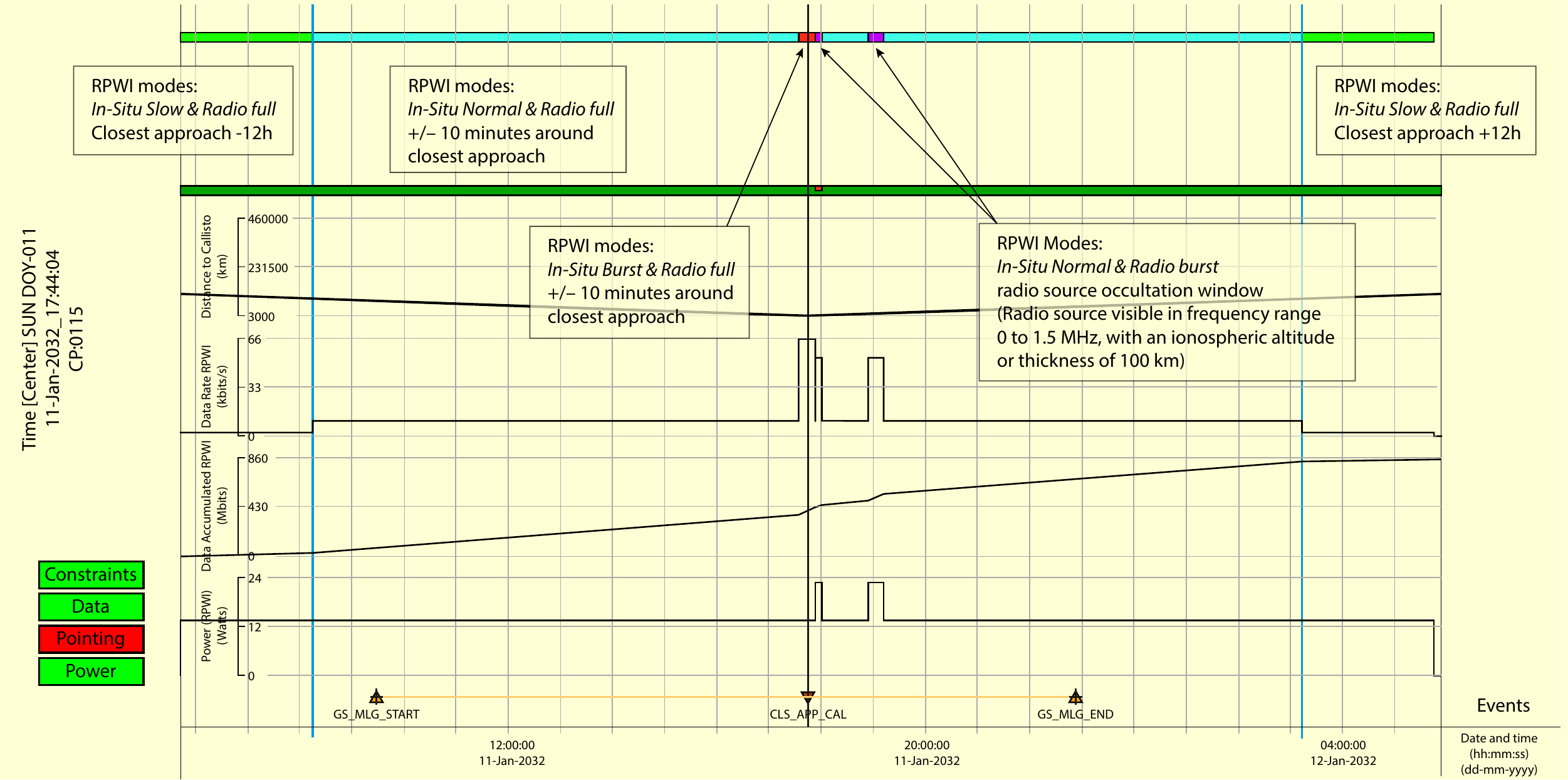}
    \caption{\revised{Screenshot of the MAPPS interface for JUICE/RPWI during 21C13. This 
    tools displays the various science segments, together with system parameters, such as: the
    distance to Callisto; the data rate produced by RPWI; the cumulated data produced by 
    RPWI; and the instantaneous power used by RPWI. The top line is colour coded 
    to reflect the different modes used by RPWI during the sequence. The annotations in the 
    boxes have been added manually for this figure, and include also the \emph{In-situ} modes, 
    which includes others RPWI subsystems, such as the Langmuir probes or the triaxial 
    magnetic search coil.}}
    \label{JUICE SOC-fig3}
\end{figure*}

\revised{Finally,} figure \ref{JUICE SOC-fig3} is a screenshot of this measurement \revised{sequence} as shown 
by the JUICE SOC Mapping and Planning Payload Science software (MAPPS), used to simulate 
the spacecraft and payload \revised{resource} status (i.e., power, data rate, on-board \revised{solid-state}
mass memory (SSMM)). The RPWI operations are planned around the CA and around ingress and 
egress ionosphere \revised{characterisation} opportunities as described in Table \ref{JUICE SOC-tab1}. 
The high-resolution in-situ measurement mode is scheduled +/- 10 minutes around the closest 
approach (CA), while the high resolution radio measurement mode is scheduled during the 
ionosphere \revised{characterisation} opportunities. High-resolution modes are the  
\revised{most} demanding in term of resources (power, data generated, stored in SSMM and to 
be downloaded (data rate)) and \revised{they} are reserved for priority scientific objectives. 
So it is crucial to be able to calculate the corresponding opportunities windows.

The JUICE mission is in development phase, with a planned launch in \revised{September} 2022. At this 
stage the JUICE mission science planning process is being exercised by analysing 
representative science scenarios similar to 21C13. In this study, the scenario analysis 
covers $\pm$12 hours around the Callisto closest approach. \revised{Figure \ref{JUICE SOC-fig3} shows 
the full interval, while Figures \ref{JUICE SOC-fig1} and \ref{JUICE SOC-fig2} are focusing on the time
interval displaying radio occultation events}. The JUICE SOC \revised{will identify} 
the same type of opportunity windows whenever a new candidate trajectory is available for JUICE.
 
The \revised{ExPRES simulation} results can also be useful for other \revised{types of measurement},
and will be made available \revised{for} JUICE SOC instrument teams. This includes the measurement
linked to the icy shell characterisation of the icy moons:
\begin{itemize}
    \item Passive radar measurement by \revised{RIME} or by RPWI: Opportunities 
    can be identified whenever any radio source is visible from the spacecraft for any 
    source type (\revised{A, B, C or D}), and per source type (to differentiate source from 
    northern and southern hemispheres) between 1 and 40 MHz (although lower frequencies are 
    better \revised{since they allow to probe deeper layers of the icy shells});
    \item Active Radar measurement by \revised{RIME}: when the spacecraft is 
    protected from Jupiter radio emission due to moon occultation for sources with frequency 
    between 9 and 11 MHz (i.e., flybys and Ganymede phase) and when the spacecraft is 
    within the RIME instrument operating range (altitude $<$1000~km).
\end{itemize}

\section{Conclusions and Perspectives} 

The Galileo radio occultations observed \revised{in} the PWS data set are well 
modelled by ExPRES simulation, with a sampling of the order of one minute. 
Discrepancies between predicted and observed ingress or egress times can be 
attributed to refraction effects, which are not included in the current modelling 
scheme. The validation on GLL/PWS data allows us to apply the same modelling to the 
JUICE mission planning, in order to support the scientific segmentation of the 
Galilean moon flybys. The ExPRES modelling will also be useful for data analysis 
of ``passive radar'' observations, since it provides the location of the Jovian 
radio sources used to probe the moon's sub-surface. 

On a technical point of view, the JUICE modelling have been done running ExPRES through 
the PADC operated UWS interface based on OPUS. This framework is fully adapted to the usage 
presented in this study. Future developments of the ExPRES code and its implementation at 
PADC will include better management of Provenance metadata
\citep{2020ivoa.spec.0411S,2021arXiv210108691S}, to enhance the scientific traceability and
reproducibility of the results.

Several means of refining the occultation modelling have been identified. The main one 
is involving ray-tracing in the Jovian system (mainly the Io Torus) and the moon's environment. 
This extra modelling step requires models of the magnetic field and plasma density 
environments in the vicinity of the studied moon, including the magnetospheric and moon 
contributions \citep[see, e.g., ][]{Modolo:2018gu}. A second order improvement may also 
be provided by the use of a more accurate magnetospheric current sheet model, which will 
refine the location of the radio sources on the low frequency end. 

\section*{Acknowledgments}
The authors acknowledge support from Observatoire de Paris--PSL, CNRS, CNES and ESA for 
funding the research. \revised{CKL’s work at DIAS is supported by the Science Foundation Ireland 
Grant 18/FRL/6199}. The authors also acknowledge support from the Europlanet 2024 Research 
Infrastructure (EPN2024RI) project, which has received funding from the European Union's 
Horizon 2020 research and innovation programme under grant agreement No 871149). They want 
to emphasise the use of community developed tools and standards, which greatly facilitated 
this study (such as Autoplot, Das2, OPUS, UWS and Cosmographia and WebGeoCalc). They thank 
PADC for providing computing and storage resources. The authors thank Ronan Modolo for fruitful 
discussions during the preparation of the manuscript. They also thank Marc Costa (from the 
NAIF SPICE team at NASA/JPL) for his very helpful feedback on configuring and using Cosmographia.

\appendix
\section{Equatorial Shadow Zone}
\label{appendix:esz}
In the equatorial region, in the innermost magnetosphere, the auroral radio sources are not 
visible, due to combination of the shape of the radio emission beaming patterns and the 
topology of the magnetic field lines bearing the radio source. This effect is named ``Equatorial 
Shadow Zone'' (ESZ). This effect has been identified at Saturn \citep{lamy_JGR_08b}, using a 
preliminary version of the ExPRES code. It has also been observed at Earth \citep[see, 
e.g.,][Figure 1]{morioka:insu-01180185}, but not explicitly described.

\begin{figure}[ht]
    \centering
    \includegraphics[width=\linewidth]{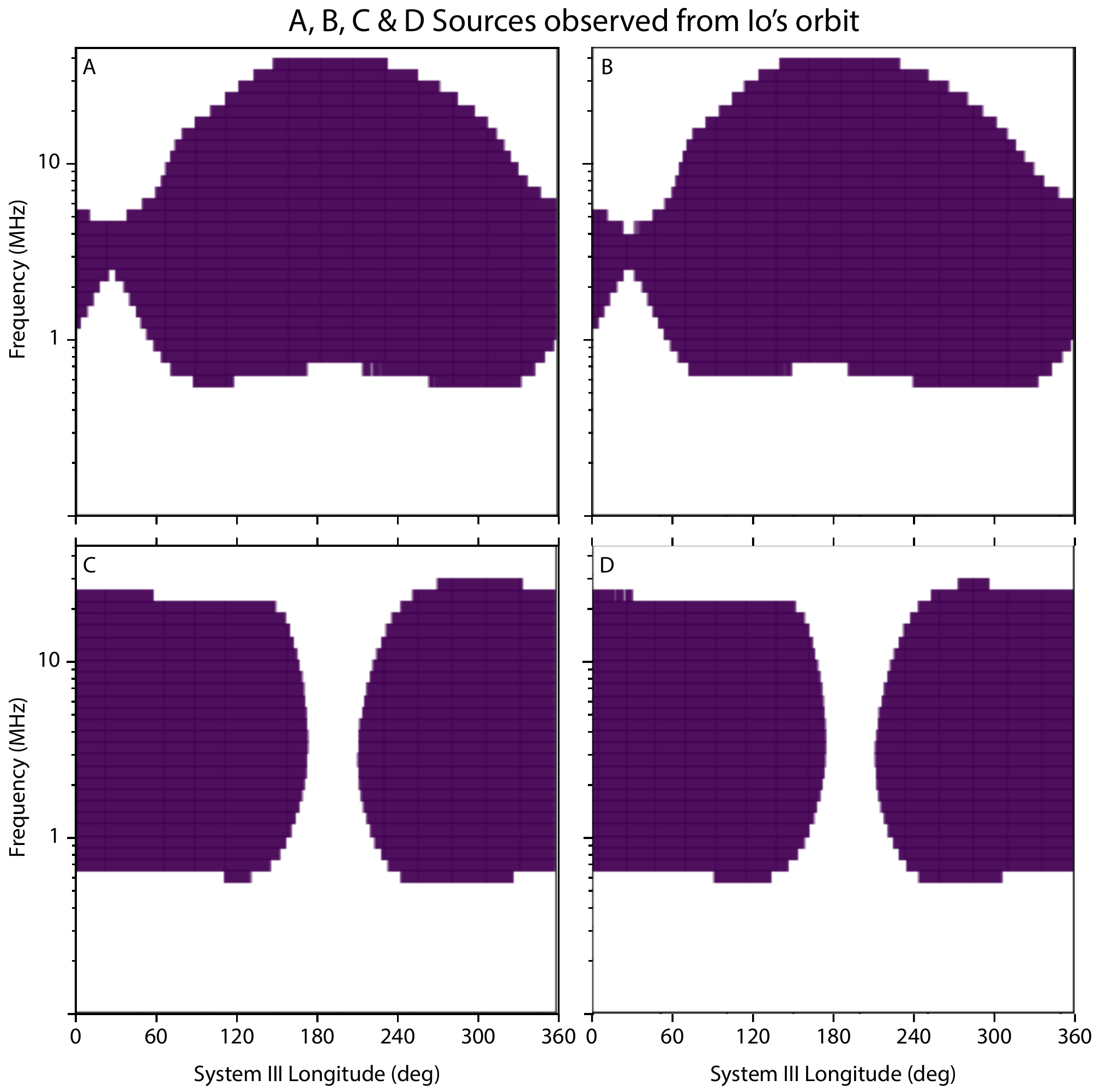}
    \caption{Jovian auroral radio source observability from Io's orbit.}
    \label{fig:esz-io}
\end{figure}

Our simulation runs show that some of the Jovian auroral radio sources are not always 
visible at the distance of Io's orbit. Figure \ref{fig:esz-io} shows the observability for 
each auroral radio source for an observer located on Io's orbit, during one rotation of 
Jupiter. The simulation shows that the Southern radio sources (namely, the C and D source) are 
not visible in the CML range 180 to 210 degrees. The Northern radio sources (namely, the A and 
B source) are observable from all CML, with a drastically reduced spectral range at a CML of 
about 25 degrees. Since the Northern and Southern ESZs do not occur at the same time, an 
observer will not experience a full dropout of radio signals, \revised{contrary} to what is 
observed at Saturn. 

\section{Radio source location uncertainty}
\label{sec:accuracy}
\begin{figure}
    \includegraphics[width=0.7\linewidth]{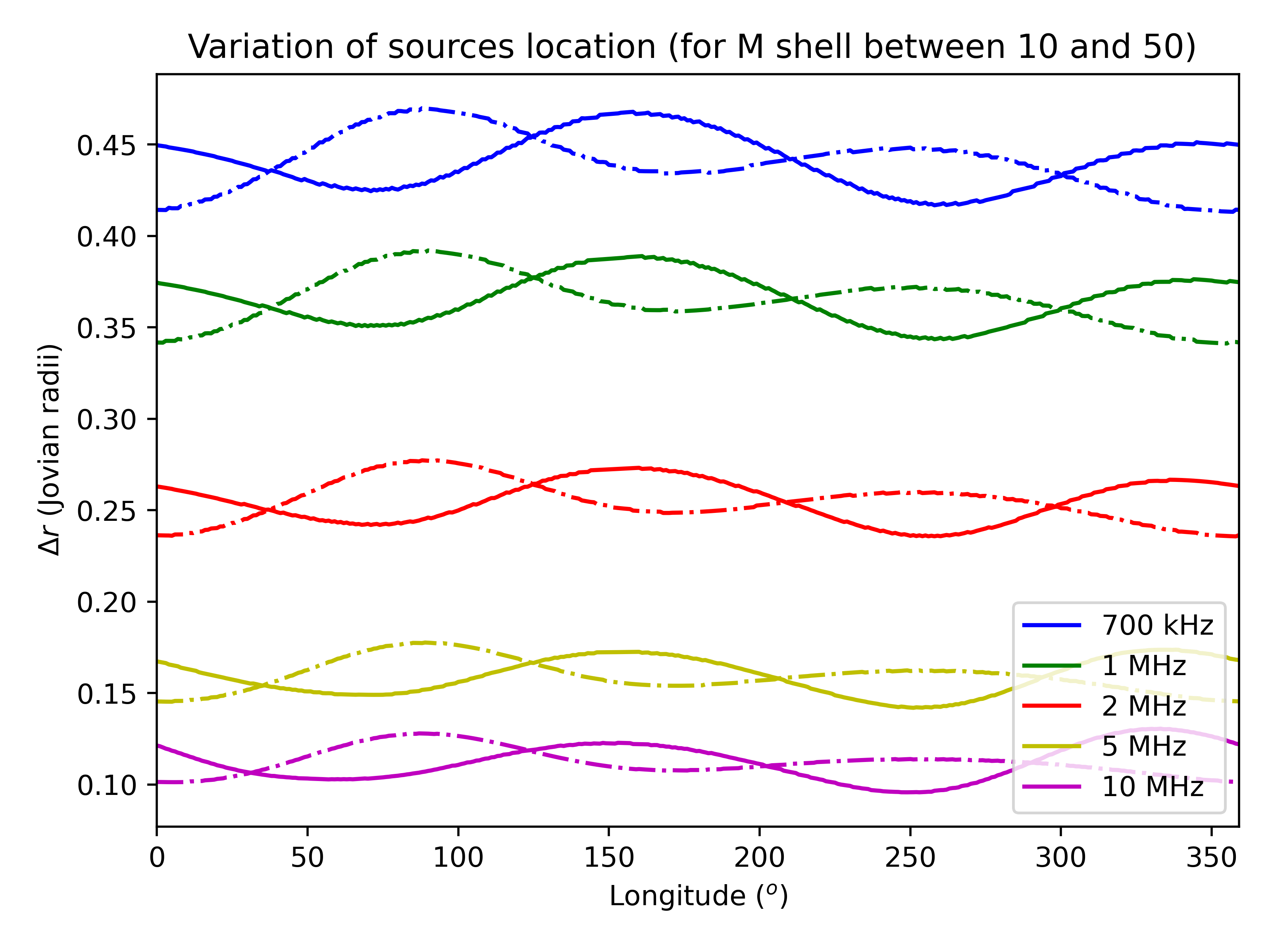}
    \caption{Radio sources location shifts (in Jovian radii) with varying M-shell (from 
    10 to 50 $R_J$), as a function of longitude, for the Northern (solid line) and Southern 
    (dotted-dashed line) hemispheres, at different frequencies: 700 kHz (blue), 1 MHz
    (green), 2 MHz (red), 5 MHz (yellow) and 10 MHz (purple).}
    \label{fig:radio-source-location}
\end{figure}

\revised{The ExPRES code provides the observable radio source locations, given a 
series of input parameters (Jovian magnetic field model, Jovian current sheet model,
electron population properties in the radio source). The main source of uncertainty 
on the radio source location is the selected M-shell of the magnetic field line 
bearing the radio sources. \citet{Louis:2019kn} identified radio sources on M-shells
between 10 and 50 Jovian radii. We thus use this range of values to evaluate 
the variation of radio source location. Figure \ref{fig:radio-source-location} shows 
the variation of the radio source locations, at five selected frequencies, for all
longitudes. We use this estimation to derive the angular uncertainty of the radio 
source location, as observed from the Galilean moons, as presented in Table 
\ref{tab:radio-source-location-accuracy}. For instance, at 10 MHz, the error 
on the source position, is below 1$^\circ$ at Europa, Ganymede and Callisto.}

\begin{table}
    \begin{tabular}{l|ccccc}
    & \multicolumn{5}{c}{Angular uncertainty (deg)} \\
    Moon & 700 kHz & 1 MHz & 2 MHz & 5 MHz & 10 MHz \\
    \hline
    Io & 4.3 & 3.5 & 2.5 & 1.6 & 1.1 \\ 
    Europa & 2.7 & 2.2 & 1.6 & 1.0 & 0.7 \\ 
    Ganymede & 1.7 & 1.4 & 1.0 & 0.6 & 0.4 \\ 
    Callisto & 1.0 & 0.8 & 0.6 & 0.3 & 0.2 
    \end{tabular}
    \caption{Angular uncertainty (in degrees) of Jovian radio sources location 
    for M-shells, as observed from each Galilean moons and for a set of 5 
    frequencies from 700 kHz to 10 MHz).}
    \label{tab:radio-source-location-accuracy}
\end{table}

\section{Evaluation of refraction effects}
\label{appendix:refraction}
\revised{The impact of propagation effects on the low frequency radio waves in the
Galilean moon environments can be evaluated using the currently available plasma
environment models (see Table \ref{tab:iono_models}). Using these models, the 
refraction index is computed after the Appleton-Hartree equation \citep{appleton}, 
which defines the refractive index for electromagnetic wave propagation in a cold 
and magnetised plasma: 
\begin{equation}
n^2(\omega) = 1- \frac{2X(1-X)}{2(1-X)-Y^2\sin^2\theta\pm\sqrt{Y^4\sin^4\theta + 
4(1-X)^2Y^2\cos^2\theta}}
\label{eq:appleton-hartree}
\end{equation} 
with $n$ the refractive index, $X = (\omega_p/\omega)^2$, $Y = \omega_c/\omega$, 
$\theta$ is the propagation angle (between the wave vector and the local magnetic 
field vector), and sign of $\pm$ is the propagation mode. The $X$ term is computed 
using the selected plasma density models. The $Y$ is evaluated using the average 
value of the Jovian magnetic field strength at each moon, that are 400 nT, 120 nT 
and 30 nT, for Europa, Ganymede and Callisto respectively. In the case of Ganymede, 
two cases have been studied, with the surface magnetic field values at the equator 
of the moon (720 nT) and at its poles (1440 nT). In all cases, the $Y^2$ value is 
negligible compared to the order of magnitude of $X$, except very close to the moon
surface. In turn, the variation with angle $\theta$ is also negligible. We thus 
only show figures for $\theta=0$ with the $+$ propagation mode and the local 
Jovian magnetic field strength, except for the case of Ganymede, where we 
considered the polar magnetic field strength.}

\revised{It is noticeable that setting $Y=0$ simplifies Equation 
\ref{eq:appleton-hartree} to:
\begin{equation}
n^2(\omega) = 1 - X = 1 - \left(\frac{\omega_p}{\omega}\right)^2
\label{eq:appleton-hartree-y-0}
\end{equation} 
So that the refractive index only depends on the plasma frequency and the radio 
wave frequency. Table \ref{tab:parameter-space} presents estimated lower plasma 
density limit and upper altitude limits for various observations frequencies at 
Europa, Ganymede and Callisto. These estimates are computed using Equation
\ref{eq:appleton-hartree-y-0} and the various environment models presented in 
this section.}

\begin{figure}
    \centering
    \includegraphics[width=0.9\linewidth]{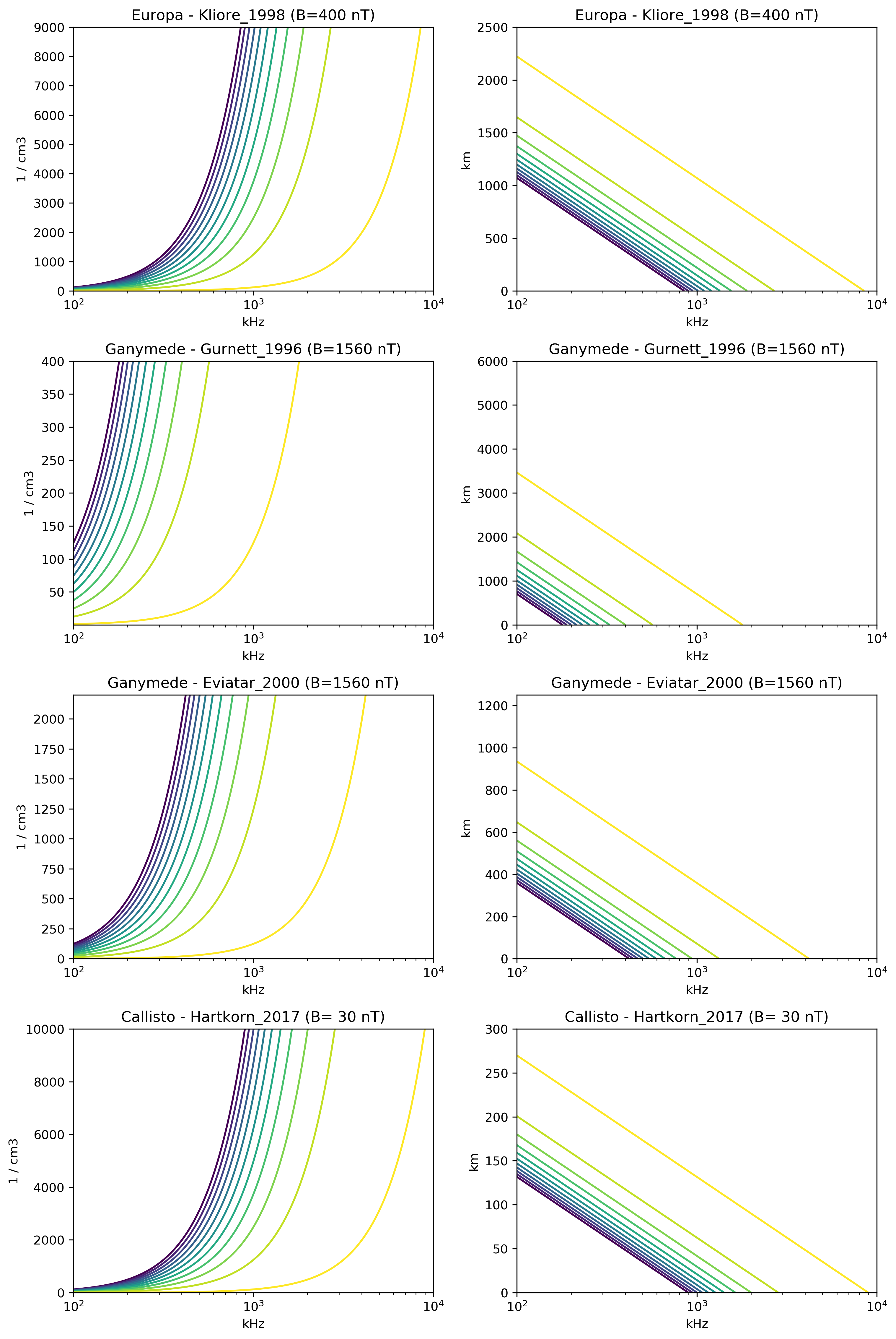}
    \caption{Contours of the refractive index values ($n^2$), for each selected environment models, 
    with 11 contour levels at 0.0, 0.1, 0.2... 0.9 and 0.99, from left (purple) to right (yellow). 
    From top to bottom: Europa with \citet{kliore:1998hp}, Ganymede with \citet{Gurnett:1996tt} 
    or \citet{Eviatar:2001ve}, and Callisto with \citet{Hartkorn:2017wx}. The left-hand side 
    columns shows the contour in the plasma density versus frequency space. The right-hand side 
    column shows the altitude versus frequency space.}
\label{fig:n2-moons}
\end{figure}

\revised{Figure \ref{fig:n2-moons} shows the squared refractive index value $n^2$ 
iso-contours in the plasma density versus frequency or altitude versus frequency 
spaces. These plots can be used to evaluate in which regimes the refraction effects 
will be observable. We consider that the $n^2=0.99$ contour provides a rough upper 
limit of this domain. Hence, for instance, with an observation frequency of 2 MHz 
at Europa, refraction effects are likely to occur up to $\sim$ 700 km above the 
surface, or at plasma densities above $\sim$ 500 cm$^{-3}$, confirming the rough 
estimate provided in Table \ref{tab:parameter-space}.}

\begin{table}
    \begin{tabular}{llll}
    Moon & $n_0\,(\textrm{cm}^{-3})$ & $h_0\,(\textrm{km})$ & Reference \\
    \hline
    Europa & 9000 & 250 & \citet{kliore:1998hp} \\
    Ganymede & 400 & 600 & \citet{Gurnett:1996tt} \\
    Ganymede & 2200 & 125 & \citet{Eviatar:2001ve} \\
    Callisto & 10000 & 30 & \citet{Hartkorn:2017wx} 
    \end{tabular}
    \caption{Plasma environment model parameters (reference density $n_0$ and scale height $h_0$) for 
    Europa, Ganymede and Callisto.}
    \label{tab:iono_models}
\end{table}

\begin{table}[ht]
    \begin{tabular}{l|l|rrrrr|l}
    \multicolumn{2}{l|}{}&\multicolumn{5}{c|}{Frequency (MHz)}&\\
    \hline
    Quantity  & Moon (Reference)&  0.7 & 1 & 2 & 5 & 10 & Unit\\
    \hline
    \hline
    Plasma Density & \emph{all}    &   60 &  125 & 500 & 3000 & 12500 & cm$^{-3}$ \\
	\hline
    Altitude  & Europa \citep{kliore:1998hp}    & 1250 & 1000 & 720 &  270 & -- & km \\
    Altitude  & Ganymede \citep{Gurnett:1996tt} & 1100 &  700 &  -- &   -- & -- & km \\
    Altitude  & Ganymede \citep{Eviatar:2001ve} &  450 &  360 & 180 &   -- & -- & km \\
    Altitude  & Callisto \citep{Hartkorn:2017wx}&  150 &  130 &  90 &   35 & -- & km 
	\end{tabular}
	\caption{Estimated parameter thresholds accessible with low frequency radio occultations. 
	The plasma density threshold is a lower limit, while the altitude threshold 
	is an upper limit. Missing values (--) correspond to situations where no refraction is 
	expected, considering no significant refraction occurs for $n^2 > 0.99$.}\label{tab:parameter-space}
\end{table}

\section{Supplementary Material Description}
\label{app:supplementary}
The material used to conduct the Galileo flybys' study are provided as a separate 
data collection \citep[available at
\url{https://doi.org/10.25935/8zff-nx36}]{https://doi.org/10.25935/8zff-nx36} 
hosted by PADC (Paris Astronomical Data Centre). For each flyby, \revised{the landing 
page provides access to: (a) the ExPRES products; (b) the Cosmographia context products; 
(c) a summary dynamic spectrum and (d) the `pov' movie}. The content of \revised{(a) and (b)} 
is described below. 

\subsection{ExPRES products}
The ExPRES data set is composed of four files: (a) an ExPRES configuration file 
(JSON format), (b) a Galileo spacecraft ephemeris data exported from WebGeoCalc 
(CSV format), (c) the output ExPRES simulation run data (CDF format), and (d) 
the observed GLL/PWS data with the superimposed occultation contours (PNG format). 
Files (a) and (b) contains the ExPRES \revised{run} parameters and the SPICE 
Kernels used for the simulation, ensuring the results are reproducible. 

\subsection{Cosmographia context products}
The Cosmographia data set is composed of a series of subdirectories, organised 
according to Cosmographia's documentation. It contains all the required configuration 
catalogue files: the SPICE catalogue file listing the kernels in use for a scene; 
the spacecraft catalogue file defining the time interval of the scene and the 
spacecraft reference frame; the modelled radio source catalogue file, derived from 
an ExPRES simulation run; the ExPRES configuration file and simulation run; and 
the scripts used to produce the movie output. Two output movies are provided, 
showing the flyby scenes, as seen from the spacecraft (`pov' labelled movie) and 
from the top of the Jovian system (`top' labelled movie).

\bibliographystyle{plainnat}
\bibliography{juice-expres}
\end{document}